\pdfoutput=1

\documentclass[journal]{IEEEtran}

\usepackage[pdftex]{graphicx}
\usepackage{amsmath}
\usepackage{array}

\ifCLASSOPTIONcompsoc
  \usepackage[caption=false,font=normalsize,labelfont=sf,textfont=sf]{subfig}
\else
  \usepackage[caption=false,font=footnotesize]{subfig}
\fi

\usepackage{url}

\usepackage{amsmath} 
\usepackage{amssymb}  

\usepackage{multirow}
\usepackage{subfig}
\usepackage[noadjust]{cite}

\usepackage{newtxtext,newtxmath}
\usepackage{bm}

\usepackage{color}

\usepackage[normalem]{ulem}
\usepackage{soul}
\usepackage{hyperref}
\usepackage{nameref}
\usepackage{xparse,xcolor}

\begin{document}

\title{Driving Skill Modeling Using Neural Networks for Performance-based Haptic Assistance}

\author{Hojin~Lee,~\IEEEmembership{Member,~IEEE},
        Hyoungkyun~Kim, 
        and~Seungmoon~Choi,~\IEEEmembership{Senior Member,~IEEE}
\thanks{H. Lee and S. Choi are with the Department of Computer Science and Engineering, Pohang University of Science and Technology, Pohang, South Korea. H. Lee is now with Haptic Intelligence Department, Max Planck Institute for Intelligent Systems, Stuttgart, Germany. S. Choi is the corresponding author. e-mail: \{hojini33, choism\}@postech.ac.kr.}
\thanks{H. Kim was with the Department of Mechanical Engineering at the same institution. He is now with the Smart Device Team, Samsung research, Seoul, South Korea. e-mail: hkyun87.kim@samsung.com}
\thanks{Manuscript received March 04, 2020; revised September 09, 2020; accepted January 28, 2021.}}

\markboth{}
{Lee \MakeLowercase{\textit{et al.}}: Driving Skill Modeling Using Neural Networks for Performance-based Haptic Assistance}

\maketitle

\begin{abstract}
    This paper addresses a data-driven framework, modeling expert driving skills for performance-based haptic assistance using neural networks (NNs).
    We have built a haptic driving training simulator to collect expert driving data and to provide proper haptic feedback.
    We establish an expert driving skill model by training NNs with the collected data.
    Then, the skill model is applied to performance-based haptic assistance to provide optimized references of the steering/pedaling movements.
    We evaluate the skill model and its application to performance-based haptic assistance in two user experiments.
    The results of the first experiment demonstrate that our skill model has appropriately captured experts' steering/pedaling skills.
    The results of the second experiment show that our performance-based haptic assistance can help novice drivers perform steering as expert drivers, but cannot assist their pedaling performance.
\end{abstract}

\begin{IEEEkeywords}
Haptic assistance, haptic shared control, artificial neural networks, driving skill, simulated driving, motor performance.
\end{IEEEkeywords}

\section{INTRODUCTION}
\label{sec:introduction}
    \IEEEPARstart{H}{aptic} assistance provides assistive feedback in the form of tactile or kinesthetic stimuli in order to facilitate the execution of or expedite the learning of motor tasks.
    In particular, kinesthetic feedback is effective in delivering mechanical momentum and moves the limbs of interest, supplying more direct, detailed, and continuous information.
    A number of studies have examined the effects of kinesthetic assistance on a variety of tasks~\cite{Heuer2015}.
    Driving is the representative area where haptic assistance has been actively studied and adopted~\cite{Gaffary2018}.
    Modern vehicles are controlled by a manual interface (steering wheel, accelerator/brake pedals, etc.) that requires complex coordinated dynamic control of the limbs.
    Therefore, while driving, the driver and the car are in a continuous loop of human-machine \emph{shared control} in which both agents constantly interact with each other and share a common goal to perform an effective, safe, and robust driving together.
    Many studies have demonstrated that haptic assistance can effectively improve the performance of basic driving maneuvers (steering~\cite{Steele2001, Griffiths2005, Forsyth2006, Saleh2013} and pedaling~\cite{Abbink2011, Mulder2011, Jamson2013}) by augmenting the shared control loop.
    This control mechanism is also called haptic shared control because the augmented information flows bidirectionally within the two agents via haptic channels at the mechanical contact of the interfaces in the loop~\cite{Omalley2006, Abbink2012}.

    \emph{Haptic guidance}, one of the effective haptic assistance methods, is the most straightforward application using haptic shared control.
    Haptic guidance provides the user with external haptic stimuli to guide the desired movement during task execution~\cite{Gillespie1998}.
    Here the haptic assistance system plays the role of a collaborator that encourages appropriate maneuvering movements and corrects the users' performance.
    Therefore, this approach is considered as a bridge to automated driving with improved performance and reduced effort~\cite{Mulder2012}.
    Automobile companies have begun to include haptic assistance in an advanced driver-assistance system for lane keeping, intelligent parking, and adaptive cruise control~\cite{Petermeijer2015}.
    Another major promising area that haptic assistance greatly contributes is \emph{skill transfer} regarding the long-term learning effects on motor skill performance~\cite{Sigrist2013}.
    Several studies have investigated the effects of various types of haptic assistance for learning driving skills in simulated environments~\cite{Marchal-Crespo2008,Marchal-Crespo2009,Lee2014}.

	\begin{figure}[!tb]
	\centering
	\includegraphics[width=\linewidth]{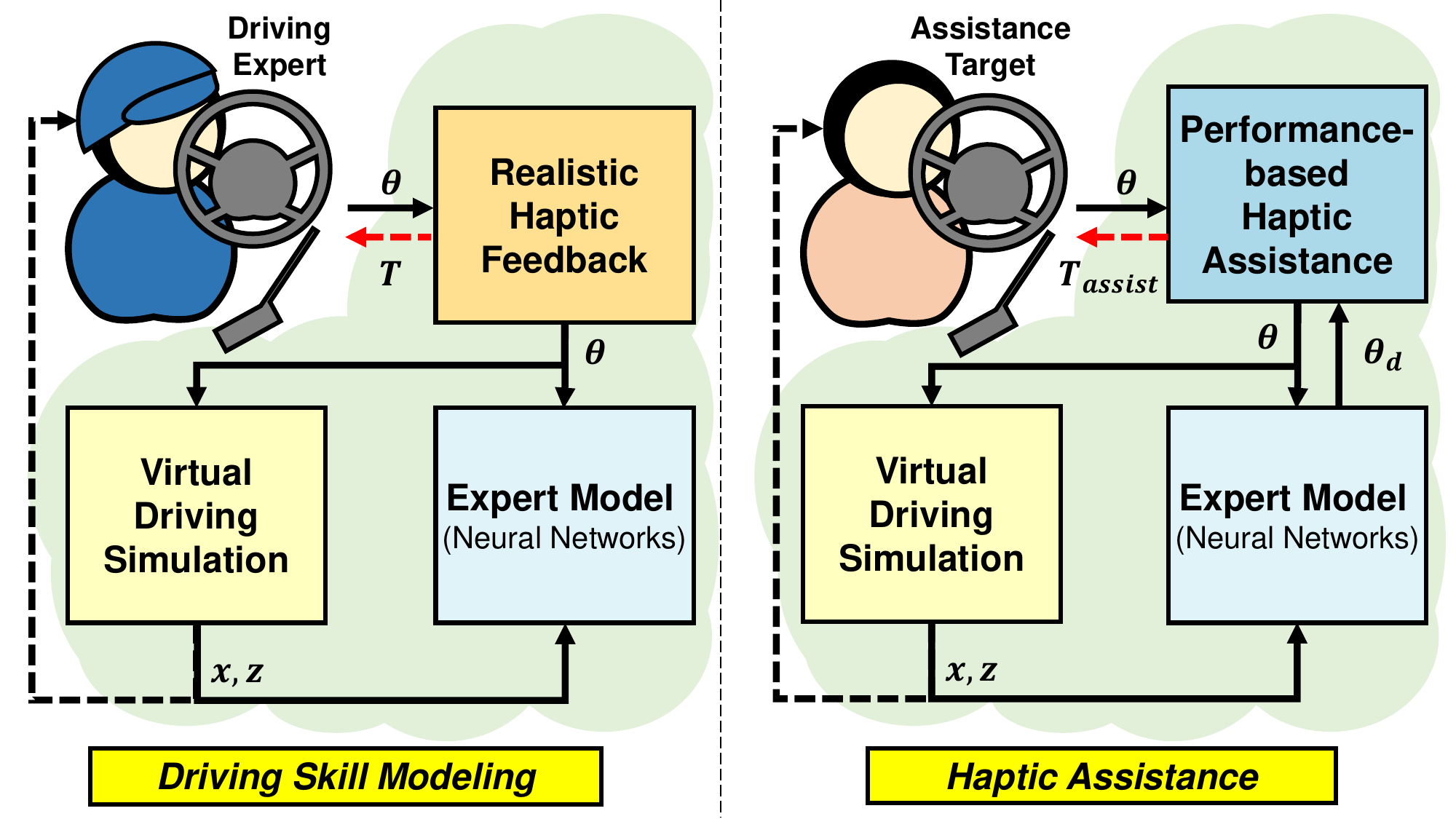}
    \caption{Simplified diagram of our framework in human-in-the-loop. Dotted lines represent the channels where a user can receive any sensory information. In particular, the red color indicates the haptic modality.}
	\label{fig:framework}
	\end{figure}

    In both areas of shared control and skill transfer, the present design is generally \emph{performance-based}; the system continuously monitors the user's current driving performance and then delivers adaptive haptic feedback.
    It requires the process of quantifying the task performance in reference to the desired driving, often called a \emph{modeling} process.
    However, the conventional modeling process essentially requires well-parameterized physical formulas and their parameters, which demand excellent prior knowledge earned from the tidy investigation about driving.
    Therefore, it is difficult to consider realistic human behaviors thoroughly, and sometimes it is simplified without behavioral characteristics.
    As a result, it is sometimes possible that the performance assessment is based on an erroneous error measure, which is likely to degrade the performance of shared control or skill transfer.

    In this paper, we address a \emph{data-driven} framework of performance-based haptic assistance, especially for driving skills~(Fig.~\ref{fig:framework}).
    First, we build an optimized black-box model for the target skill and then use the skill model for performance-based haptic assistance.
    We record how experienced drivers perform driving (without haptic assistance) in realistic virtual environments, and then train the adequate continuous model of relevant variables from the collected data.
    The data-driven approach is being popularly considered and adopted as a model of driving skills, sometimes with advanced machine-learning techniques~\cite{Aksjonov2018, Abou2020}.
    In this regard, we use shallow but cost-effective (artificial) neural networks (NNs) in order to provide real-time haptic feedback.
    Some researchers have suggested the feasibility of NN models in accounting for human driving skills of steering control~\cite{Nechyba1996,Nechyba1997,Nechyba1998,Lin2005,Garimella2017}.
    In particular, Nechyba and Xu used NNs to model human driving strategies from data collected in simplified driving simulations using a mouse interface~\cite{Nechyba1996,Nechyba1997,Nechyba1998}.
    Their NN models can produce predictive trajectories of steering and pedaling, based on individual data.
    However, because their research did not include realistic driving hardware, the usefulness of their behavioral model and its applicability to haptic assistance require further verification.

\begin{figure}[!tb]
\centering
	\includegraphics[width=0.8\linewidth]{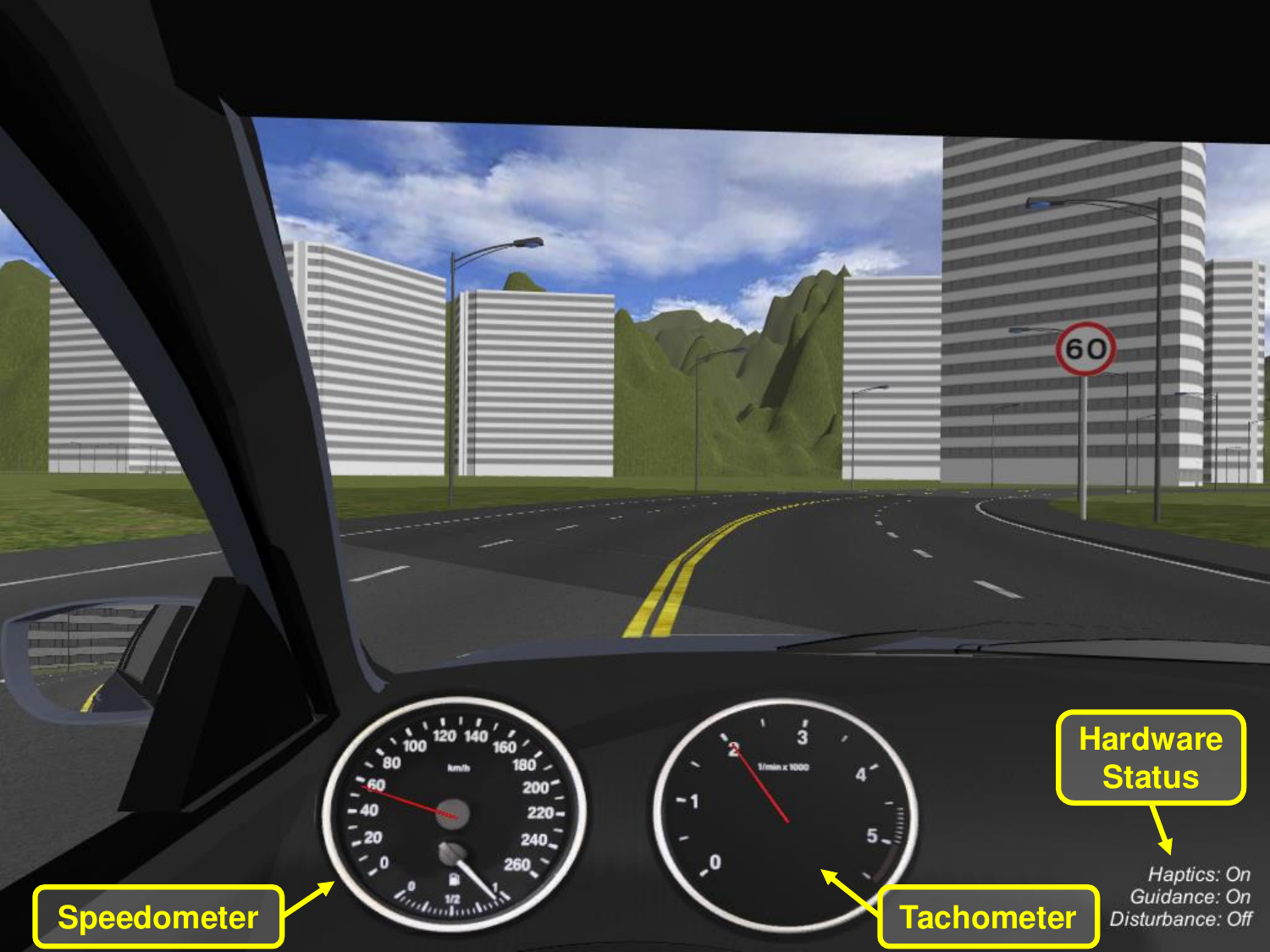}
	\caption{Visual illustration of a driving simulation scene.}
	\label{fig:screenshot}
\end{figure}

    Consequently, the NN models trained with expert driving data may compute and provide optimized device trajectories, as an expert driving skill model.
    The expert skill model is then integrated into performance-based haptic assistance to provide appropriate haptic feedback as a reference of expert actions in steering and pedaling.
    We assess the adequacy of the expert skill model and its application to the framework, in two user experiments.
    Experiment~I validates the skill model in terms of generalizability to complex environments and expressiveness of expert driving skills.
    In Experiment~II, we test if our performance-based haptic assistance can actually enhance the driving performance of novice drivers driving when the skill models are used as a reference.
    To our knowledge, novel contributions of this work lie with the introduction of data-driven skill models that represent human expert skills for haptic assistance.

\section{SIMULATOR}
\label{sec:simulation}

\begin{figure}[!tb]
\centering
	\includegraphics[width=0.85\linewidth]{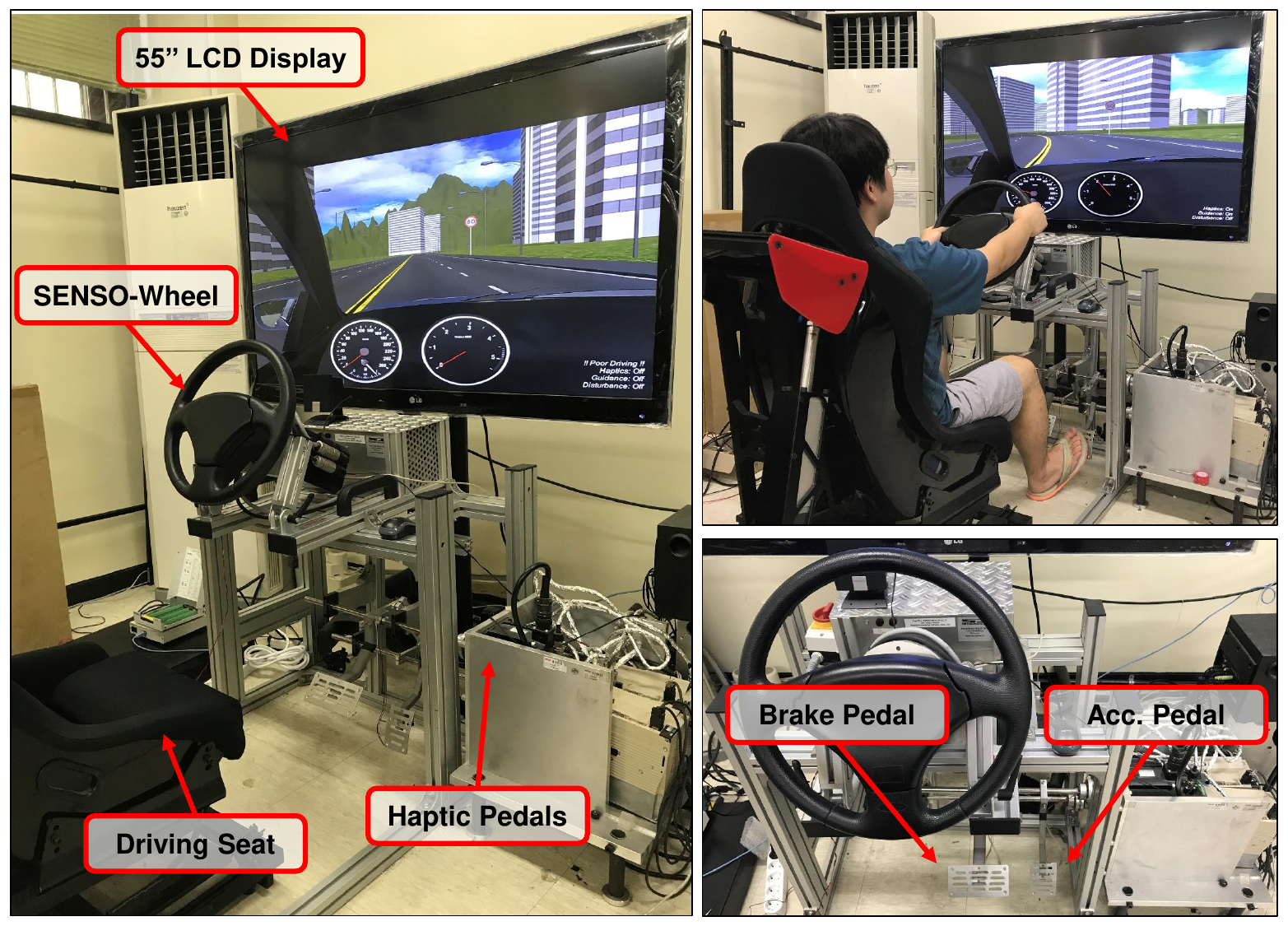}
	\caption{Haptic driving training simulator.}
	\label{fig:simulator}
\end{figure}

	We developed a driving simulator (Fig.~\ref{fig:screenshot} and \ref{fig:simulator}) for data acquisition and haptic assistance.
    For data acquisition, the simulator records the driving data of users.
    It provides realistic driving environments, including torque feedback to the steering wheel and pedals.
    For haptic assistance, the simulator generates torque feedback assisting driving skills.
    It renders audiovisual driving environments using Vehicle Physics Pro (VPP) \cite{VPP}, a commercial vehicle physics engine running in Unity~5 (update rate 50\,Hz).
    For car dynamics, a specific vehicle (Genesis, Hyundai Motors) is chosen to determine the physical parameters of VPP, such as the mass (1900\,kg), dimension (1.8\,W\,$\times$\,4.6\,D\,$\times$\,1.5\,H, in meters), steering ratio (12.0:1), gear ratios, and engine power curves~\cite{AutomobileCatalog}.

\subsection{Hardware}
\label{sec:simulation:hardware}

	The simulator consists of a large visual display, a steering wheel, an accelerator pedal, and a brake pedal~(Fig.~\ref{fig:simulator}).
	All devices are fastened to an aluminum frame to imitate a real driving seat.
	We use a 55-inch LCD (55LW6500, LG Electronics), and the distance from the display to the seat is approximately 1.2\,m for a comfortable field of view of 60\,$^\circ$.
	The simulator uses a commercial steering wheel (SENSO-Wheel SD-LC, SensoDrive) to provide high-fidelity torque feedback.
    The maximum instantaneous and continuous torques are 16.58\,Nm and 7.5\,Nm, respectively.

	We custom-designed the accelerator and brake pedals.
    Two sets of AC servo motor (SGMGV-20A, Yaskawa Electrics) and servo pack (SGDV-18011A, Yaskawa Electrics) are used to provide independent torque feedback.
	The devices and PC communicate using a MechatroLink-II network control board (PCI-R1604-MLII, Ajinextek).
    The maximum instantaneous and continuous torques of the two motors are 27.8\,Nm and 10\,Nm, respectively.
	For compact housing, both motors should be mounted on the same side, maintaining the alignment of the two rotation axes of the pedals.
    For this reason, while the accelerator pedal is directly connected to one motor with a coupler, the brake pedal is connected to the other motor through a four bar mechanism.
	The steering wheel and the pedals are controlled at a sampling rate of 800\,Hz.

\subsection{Torque Feedback}
\label{sec:simulation:torque}

    The steering wheel angle $\theta_{s}$ is between $\theta_{s, min}$ (-459\,$^\circ$) and $\theta_{s, max}$ (459\,$^\circ$) with the built-in simulation of mechanical end stops in both clockwise and counterclockwise directions.
    To provide useful information about the road and vehicle status, the simulated steering torque $T_{s}$ is implemented such that it is similar to the real torque transmitted from the driving shaft:
\begin{align}
\label{eq:Ts}
	T_{s} & = T_{s, align} + T_{s, damping} + T_{friction},
\end{align}
	where $T_{s, align}$ is the self-alignment torque, and $T_{s, damping}$ and $T_{friction}$ are the viscous and Coulomb frictions computed from the car dynamics, respectively.
	In four-wheel drive, the steering reactive torque can be estimated as follows~\cite{hiraoka2008}:
\begin{align}
\label{eq:Talign}
	T_{s, align} \approx G_{shaft} \cdot \frac{1}{2} \left( F_{fl} + F_{fr} \right),
\end{align}
	where $F_{fl}$ and $F_{fr}$ are the lateral forces applied to the left and right front wheels obtained from VPP.
    $G_{shaft}$ is the simplified gain of torque transmission from the shaft.
    $T_{s, damping}=-D_{s} \dot{\theta}_{s}$, and $T_{s, friction}$ is a constant, both in the opposite direction of the steering wheel rotation.
	From \eqref{eq:Talign}, a user can perceive driving-like sensations on the road with respect to the direction and velocity of the virtual vehicle.
	
	The haptic pedals are controlled using a spring-damper impedance control scheme.
    If the accelerator angle $\theta_{a}$ is between $\theta_{a, min}=0^\circ$ and $\theta_{a, max}=10^\circ$, it is normalized and sent to the throttle value of the virtual car engine in VPP.
	The torque to the accelerator is computed as follows:
\begin{align}
\label{eq:Ta}
	T_{a} & = T_{a, spring} + T_{a, max} + T_{a, damping} + g \left( \theta_{a} \right),
\end{align}
	where $T_{a, damping}=-D_{a} \dot{\theta}_{a}$ is the virtual damping torque and $g \left( \cdot \right)$ is for gravity compensation.
	The spring-like restoring torque $T_{a, spring}$ is determined by
\begin{align}
\label{eq:Taspring}
	T_{a, spring} & = -K_{a} \left( \theta_{a} - \theta_{a, 0} \right ),
\end{align}
	where $K_{a}$ is the virtual spring coefficient, and $\theta_{a, 0} = \theta_{a, min} - 5^\circ = -5^\circ$ is the initial position of the accelerator pedal.
    $T_{a, spring}$ pushes the driver's right foot upward to deliver information about how much s/he is pressing the pedal from $\theta_{a, 0}$.
    $T_{a, max}$ is a unilateral feedback term to provide information regarding the maximum angle such that
\begin{align}
\label{eq:Tamax}
	T_{a, max} & =
	\begin{cases}
		0 & \: \text{if } \theta_{a} < \theta_{a, max} \\
		-K_{a, max} \left( \theta_{a} - \theta_{a, max} \right) & \: \text{if } \theta_{a} \geq \theta_{a, max}
	\end{cases}
    .
\end{align}
	$T_{a, max}$ enables the driver to perceive the virtual endpoint at $\theta_{a, max}=10^\circ$.
    We used $K_{a, max} = 10 K_{a}$.

	The torque to the brake pedal, $T_{b}$, is computed similarly for the brake angle $\theta_b$.
    The only difference was that the maximum brake angle $\theta_{b, max}=5^\circ$.

\begin{table}[!tb]
\centering
\caption{Constant values for driving torque feedback}
\label{table:param}
\begin{tabular}{c|c||c|c}
\multicolumn{2}{c||}{Steering Wheel} & \multicolumn{2}{c}{Accelerator/Brake Pedals} \\ \hline
$G_{shaft}$ (m) & 0.75 & $K_{a}$, $K_{b}$ (N$\cdot$m/degree) & 0.2 \\
$D_{s}$ (N$\cdot$m$\cdot$s/degree) & 0.002 & $D_{a}$, $D_{b}$ (N$\cdot$m$\cdot$s/degree) & 0.001 \\
$T_{friction}$ (N$\cdot$m) & 0.1 &  &
\end{tabular}
\end{table}

	We carefully tuned all of the other parameters to achieve realistic experiences, and their values are specified in Table~\ref{table:param}.

 \section{SKILL MODELING USING NEURAL NETWORKS}
\label{sec:modeling}

    Performance-based haptic assistance requires an optimal (desired) behavior as the ground truth for error computation~\cite{Marchal-Crespo2009Review}.
    In our case, the task performance can be represented by the position error $\bm{e_{\theta}} = \bm{\theta} - \bm{\theta}_{d}$, for the current position (of the steering wheel and the brake pedal) $\bm{\theta} = \left[ \theta_{s} \: \theta_{a} \right]^{T}$ and the desired position $\bm{\theta}_{d} = \left[ \theta_{s, d} \: \theta_{a, d} \right]^{T}$.
    In our approach, the desired position $\bm{\theta}_{d}$ is generated by two respective NNs for $\theta_{s, d}$ and $\theta_{a, d}$.
    This is the most distinctive feature compared to prior schemes for performance-based haptic assistance.
    The two NNs are trained with prerecorded trajectories of successful driving runs by experienced drivers, which represent optimized steering and pedaling actions.

\subsection{Neural Network Structure}
\label{sec:modeling:structure}

	In \cite{Nechyba1995}, the dynamic nature of human control strategy is abstracted into a static mapping between input and output using feed-forward neural networks.
	In fact, a dynamic system can be approximated using difference equations~\cite{Narendra1990}, such that
\begin{align}
\label{equ:nn}
	\begin{aligned}
		\bm{u} \left[ k+\tau \right] = f [  & \bm{u} \left[ k \right], \bm{u} \left[ k-\tau \right], \cdots, \bm{u} \left[ k- \left( D_{\bm{u}}-1 \right) \tau \right], \\
			&\bm{x} \left[ k \right], \bm{x} \left[ k-\tau \right], \cdots, \bm{x} \left[ k- \left( D_{\bm{x}}-1 \right) \tau \right], \\
			&\bm{z} \left[ k \right], \bm{z} \left[ k-\tau \right], \cdots, \bm{z} \left[ k- \left( D_{\bm{z}}-1 \right) \tau \right]],
	\end{aligned}
\end{align}
    where $f \left[ \cdot \right]$ represents a nonlinear function, $\bm{u} \left[ k \right]$ is the control vector, $\bm{x} \left[ k \right]$ is the system state vector, and $\bm{z} \left[ k \right]$ is a vector that describes exogenous environmental features, all at time $k$.
	Then, \eqref{equ:nn} can be rewritten as
\begin{align}
\label{eq:nnAbbreviation}
	\begin{aligned}
		\bm{u} \left[ k+\tau \right] = f \left[ \bar{\bm{u}} \left[ k \right], \bar{\bm{x}} \left[ k \right], \bar{\bm{z}} \left[ k \right] \right],
	\end{aligned}
\end{align}
    where $\bar{\bm{m}} \left[ k \right] = \left[ \bm{m} \left[ k \right], \bm{m} \left[ k - \tau \right], \cdots, \bm{m} \left[ k - \left( D_{\bm{m}} - 1 \right) \tau \right] \right]^{T}$ for an arbitrary vector $\bm{m}$.

    We use a neural network to find $f$ providing the estimate $\hat{\bm{u}} \left[ k \right] = f \left[ \bar{\bm{u}} \left[ k \right], \bar{\bm{x}} \left[ k \right], \bar{\bm{z}} \left[ k \right] \right]$.
    The network is trained on the input-output data by minimizing the cost function
\begin{align}
\label{eq:C}
    \begin{aligned}
        C & = RMS \left( \tilde{\bm{d}} \right), \\
        \bm{d} & = \hat{\bm{u}} \left[ k \right] - \bm{u} \left[ k+\tau \right],
    \end{aligned}
\end{align}
    where $\tilde{\bm{m}}$ is a time series of $\bm{m}$, and $RMS \left( \tilde{\bm{m}} \right)$ computes the root mean square of all data in $\tilde{\bm{m}}$.
    In results, the output vector $\hat{\bm{u}} \left[ k \right]$ from the neural network estimates $\bm{u} \left[ k + \tau \right]$ (after $\tau$-samples) from the current and previous states of $\bar{\bm{u}} \left[ k \right]$, $\bar{\bm{x}} \left[ k \right]$, and $\bar{\bm{z}} \left[ k \right]$.

\subsection{Data Acquisition}
\label{sec:modeling:acquisition}

\begin{figure}[!tb]
\centering
	\subfloat[]
	{
      \includegraphics[width=0.27\linewidth]{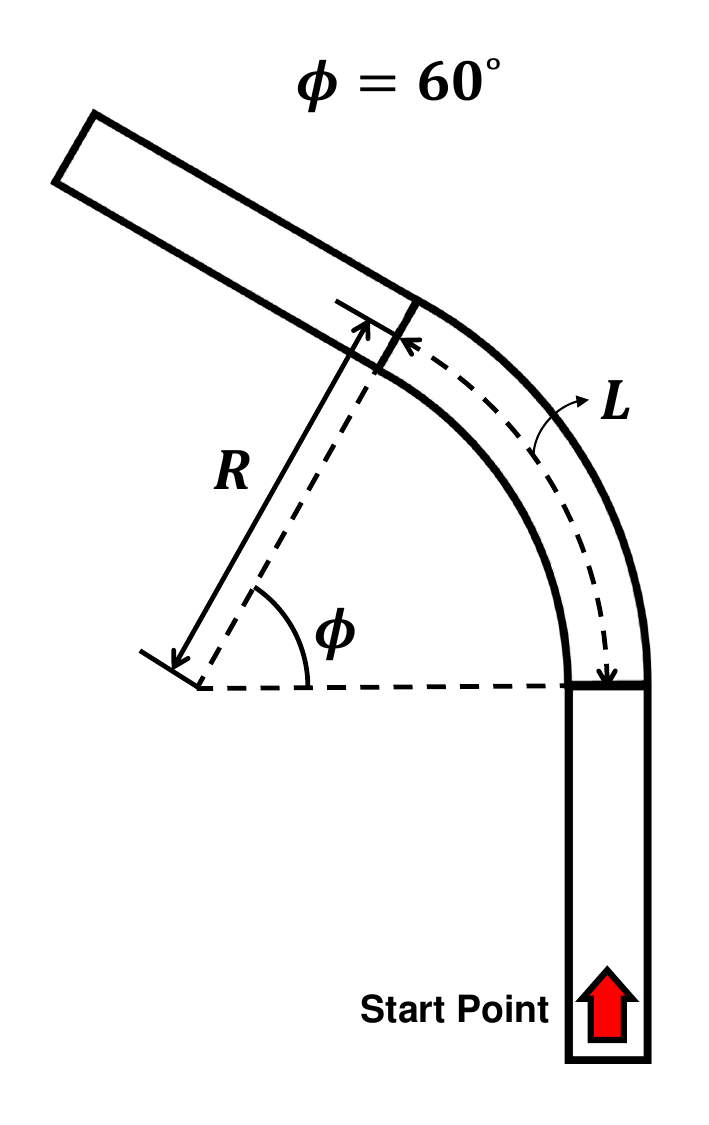}
	\label{fig:compare:e1sb}
	}
	\subfloat[]
	{
      \includegraphics[width=0.5\linewidth]{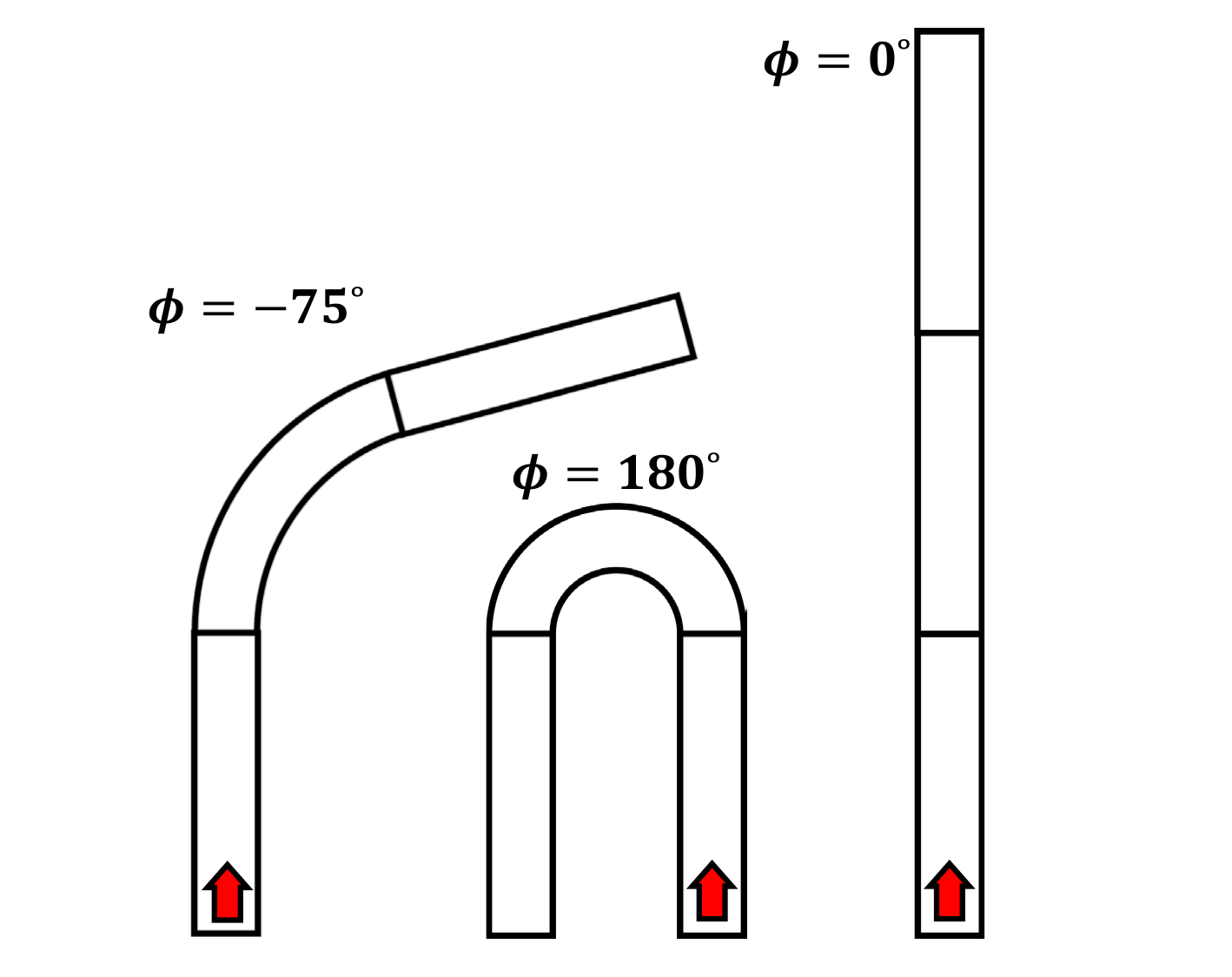}
	\label{fig:compare:e1sd}
	}
	\caption{Design of driving paths. (a) Variable definitions. (b) Three simple driving path examples in three separated segments, the $\phi$'s of the second curve segment of which are -75$^\circ$, 180$^\circ$, and 0$^\circ$, respectively.}
	\label{fig:driving_paths}
\end{figure}

	We designed 25 two-lane paths to collect driving trajectories and other important variables for skill modeling.
    The width of each lane is determined by the real standard of 3.8\,  m~\cite{HCM2010} which is enough to maintain the free-flow speed of the vehicle~\cite{Neudorff2016}.
    Each path consists of three segments with a total length of 600\,m.
	The first and the third are 200-m straight segments.
	The second is a curve with curvature $\kappa= 1/R = |\phi|/L$, where $R$ is the radius, $L$ is the arc length, and $\phi$ is the angle in radian (Fig.~\ref{fig:compare:e1sb}).
	The value $L$ of the second segment is 200\,m, but each path has varying $\phi$ from -180$^\circ$ to 180$^\circ$ in 15$^\circ$-step (Fig.~\ref{fig:compare:e1sd}).
	$\phi=0^\circ$ results in a 600-m-long straight path.

    Five experienced drivers ($\mathsf{E_1}$--$\mathsf{E_5}$; all males; age 25--51 years, M~37.6, SD~10.8; driving experience 5--30 years, M~15.2, SD~10.3) participated in the data acquisition.
    They were instructed to complete driving while staying only in the first lane of the path and maintaining 60\,km/h velocity on the speedometer.	
    Each trial took about 36--40\,s, and each driver completed six trials for each path (150\,trials per driver).
    All human experiments in this paper (data acquisition, Experiment~I, and Experiment~II) were approved by the Institutional Review Board at the authors' institution (PIRB-2017-E066).

\subsection{Neural Network Design and Implementation}
\label{sec:modeling:design}

\begin{figure}[!tb]
\centering
	\includegraphics[width=0.65\linewidth]{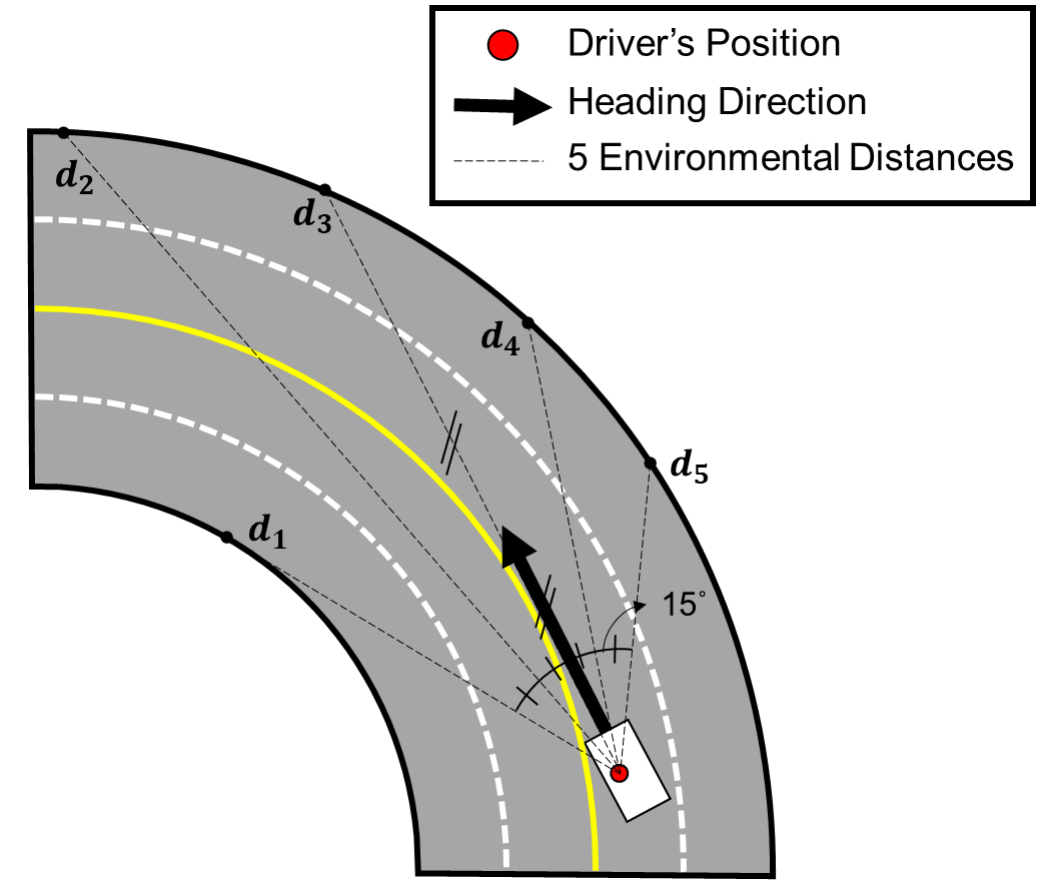}
	\caption{Five distances from the driver's perspective. Each $d_i$ means the two-dimensional Euclidean distance from the driver's position to the road boundary in five directions.}
	\label{fig:environments}
\end{figure}

	During the data collection, the experienced drivers did not use the brake pedal for lane keeping and speed control.
    So we exclude $\theta_{b}$ from the control vector.
	In addition, we do not consider the interdependence in control between the steering wheel and the accelerator pedal and train separate neural networks for each.
    This allows us to use more compact networks with accurate modeling results.
    Hence, in the model for steering, $\bm{u} = \theta_{s}$, and in the model for pedaling, $\bm{u} = \theta_{a}$.
    For the vehicle state, we use $\bm{x}= \left[ v \,\, \omega \,\, r \right]^{T}$, where $v$ is the longitudinal velocity\,(m/s), $\omega$ is the angular velocity\,(degree/s), and $r$ is the engine revolutions per minute (RPM) of the virtual car.

    The environmental features are defined by using $d_i \,( i = 1, \cdots,\, 5$; Fig.~\ref{fig:environments}).
    Total five directions in relative angles of -30$^\circ$, -15$^\circ$, 0$^\circ$, 15$^\circ$, and 30$^\circ$ to the frontal direction are selected.
	The angles are determined considering the driver's vision capability based on the field of view ($60^\circ$) within the simulated vehicle.
	The maximum value of $d_{i}$ is set to 60\,m.
    Then, the environmental feature vector $\bm{z}= \left[ z_1 \,\, z_2 \,\, z_3 \,\, z_4 \,\, z_5 \right]^{T}$, where
\begin{align}
\label{eq:zi}
	z_{i} = \frac{1}{ 1 + d_{i}}.
\end{align}
    $z_i$ represents the future hazard of collision in the $i$-th direction.

	The two NNs, $f_{s}$ and $f_{a}$, for the steering wheel and the accelerator pedal are trained with the prerecorded input-output data as follows:
\begin{align}
	\theta_{s} \left[ k + \tau \right] = f_{s} \left[ \bar{\theta}_{s} \left[ k \right], \bar{\bm{x}} \left[ k \right], \bar{\bm{z}} \left[ k \right] \right], \label{eq:NNB1}
\\
	\theta_{a} \left[ k + \tau \right] = f_{a} \left[ \bar{\theta}_{a} \left[ k \right], \bar{\bm{x}} \left[ k \right], \bar{\bm{z}} \left[ k \right] \right]. \label{eq:NNB2}
\end{align}
    The trained outputs $\hat{\theta}_s$ and $\hat{\theta}_a$ then approximate the control action that the experienced drivers would make after $\tau$ given the current and previous states of $\theta_{s}$, $\theta_{a}$, $\bm{x}$, and $\bm{z}$.

	Considering that the human motion bandwidth is less than 5\,Hz~\cite{Brooks1990}, we use the same constants: $\tau=10$ and $D_{\bm{u}}=D_{\bm{x}}=D_{\bm{z}}=5$.
	Then, the NNs simulate 0.2-s future values of execution from the five current and previous variables in 50 Hz.
	Before training, all input-output vectors, $\bm{u}$, $\bm{x}$, and $\bm{z}$ are normalized.

\subsection{Training Results}
\label{sec:modeling:results}

	We trained all NNs using MATLAB (R2017a, MathWorks).
	Specifically, we use gradient descent backpropagation with an adaptive learning rate and a transfer function of the hyperbolic tangent sigmoid.
	The initial learning rate is 0.5.
	Each NN consists of four hidden layers with 32, 24, 16, and 8 nodes.
	The input-output data of all the experienced drivers are pooled for training.
    The data are partitioned into training, validation, and test sets in the proportions of 70\,\%, 15\,\%, and 15\,\%, respectively.
    Training is terminated if the cost function $C < \delta$ (see \eqref{eq:C}), where $\delta_{s} = 1.0\,\%$ for $\theta_{s}$ and $\delta_{a} = 4.5\,\%$ for $\theta_{a}$.

\section{Performance Measures}
\label{sec:measure}

    In this section, we introduce feasible metrics to analyze and evaluate driving performance.

\subsection{Modeling Performance}
\label{sec:measure:performance}

    The modeling performance of the NN models can be evaluated by comparing original steering/pedaling trajectories ($\theta_{s}$ and $\theta_{a}$) and predicted output trajectories ($\hat{\theta}_{s}$ and $\hat{\theta}_{a}$), which are generated by the NN models from the original trajectories.
    The following error metrics indicate the error differences between predicted driving actions estimated by the NNs and a driver's real driving actions at a discrete time sample $k$:
\begin{align}
\label{eq:eps}
	e_{s, p} \left[ k \right] &= \hat{\theta}_{s} \left[ k \right] - \theta_{s} \left[ k + \tau \right], \\
	e_{a, p} \left[ k \right] &= \hat{\theta}_{a} \left[ k \right] - \theta_{a} \left[ k + \tau \right].
\end{align}
	The normalized RMSEs, $\bar{E}_{s, p}$ and $\bar{E}_{a, p}$, for each individual driving data are defined as:
\begin{align}
\label{eq:epn}
	\bar{E}_{s, p} = \frac{E_{s, p}}{\theta_{s, M} - \theta_{s, m}} = \frac{ RMS \left( \tilde{e}_{s, p} \right)}{\theta_{s, M} - \theta_{s, m}}, \\
	\bar{E}_{a, p} = \frac{E_{a, p}}{\theta_{a, M} - \theta_{a, m}} = \frac{ RMS \left( \tilde{e}_{a, p} \right)}{\theta_{a, M} - \theta_{a, m}},
\end{align}
    where $\theta_{s, M}$, $\theta_{s, m}$, $\theta_{a, M}$, and $\theta_{a, m}$ are the maximum and minimum device angles obtained from the data of the experienced drivers used for the NN modeling (these values were also used for the training data normalization in Section~\ref{sec:modeling:design}).
	$\bar{E}_{s, p}$ and $\bar{E}_{a, p}$ quantify the similarity of the participant's driving skills to those of the experienced drivers captured in the NN models.

\begin{figure} [!tb]
\centering
	\includegraphics[width=0.85\linewidth]{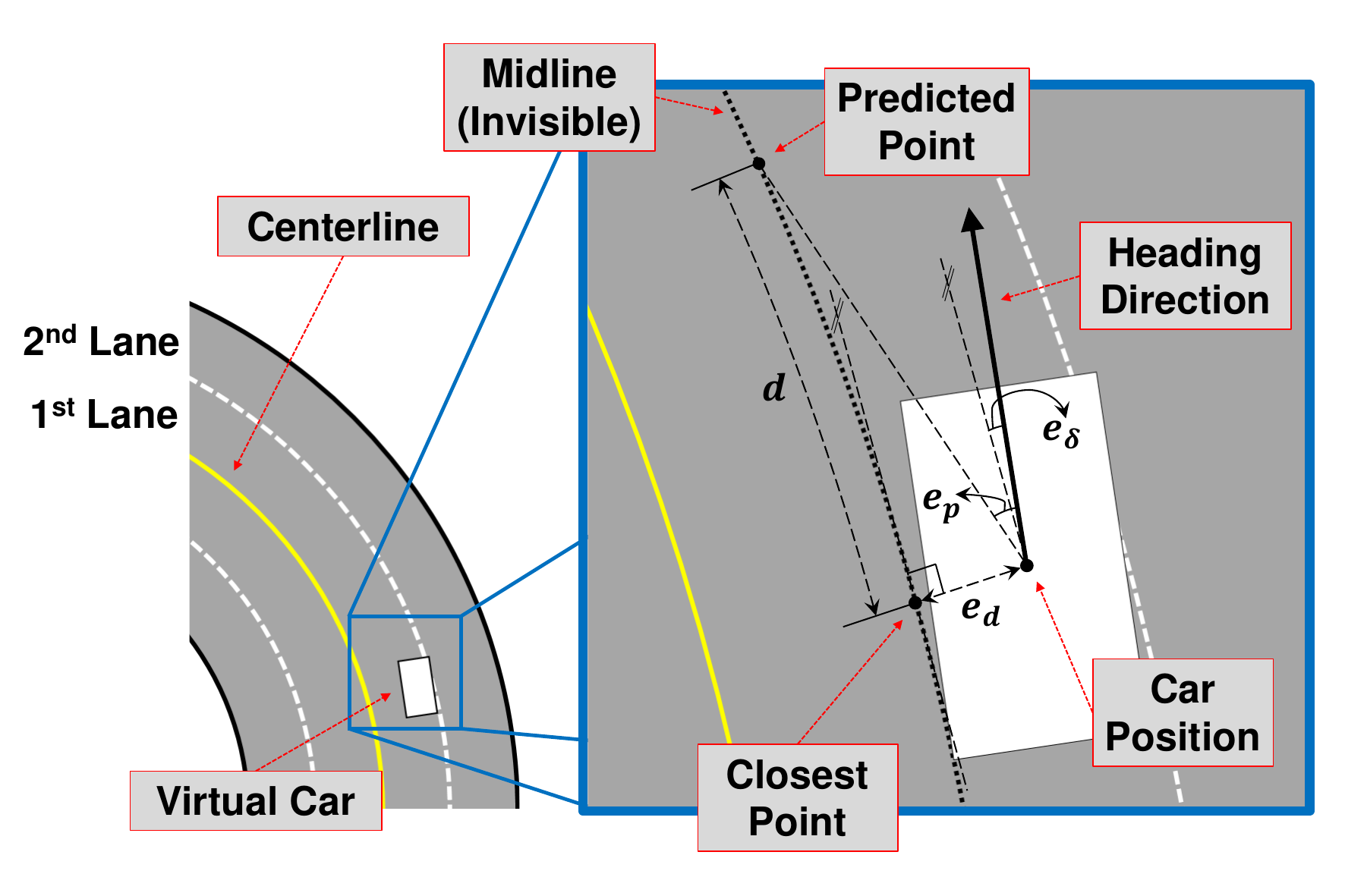}
	\caption{Driving errors ($e_{d}$, $e_{\delta}$, and $e_{p}$) used in Experiment~I and~II.
	In our simulation, $d = v \Delta t$, where $v$ is the current vehicle speed and $\Delta t = 1$\,s is the look-ahead time.}
	\label{fig:exp_error_diagram}
\end{figure}

\subsection{Trajectory-based Skill Performance}
\label{sec:measure:objective}

    The driving skill of each participant is broken down into steering and pedaling performance.
    The steering performance is evaluated by a distance error $e_{d}$ and an angular error $e_{\delta}$ (Fig.~\ref{fig:exp_error_diagram}).
	$e_{d}$ is the lateral distance between the virtual car position and the closest point on the (invisible) midline of the first lane.
	$e_{\delta}$ is the angle between the car heading and the road frontal direction at the closest point on the midline.
	We use $E_{d} = RMS \left( \tilde{e}_{d} \right)$ and $E_{\delta} = RMS \left( \tilde{e}_{\delta} \right)$ as indicators of the steering performance.

	For the pedaling performance, we define a vehicle velocity error as $e_{v} \left[k \right] = v \left[ k \right] - v_{d}$, where $v_{d}=62.64$\,km/h.
    In our simulator, the target speed of 60\,km/h corresponds to the actual speed of $v_{d}$ when the speedometer needle reaches 60\,km/h from the driver's viewpoint.
    So $E_{v}=RMS \left( \tilde{e}_{v} \right)$ is a measure for the pedaling performance.
    $E_{v}$ is computed using only the velocity samples obtained after the vehicle speed first reaches $v_d$.
    In addition, as a measure for the pedaling efficiency, we compute $\Omega_{a} = RMS \left( \tilde{\omega}_{a} \right)$, where $\omega_{a} \left[ k \right] = |\dot{\theta}_{a} \left[ k \right] |$, focusing on the pedaling speed.
	$\Omega_{a}$ increases if the participant operates the pedal more abruptly.

\subsection{Driving Behavior Performance}
\label{sec:measure:humanfactor}

    Although the trajectory-based metrics can indicate average driving performance, analysis relying on them may overlook drivers' behavioral characteristics closely related to the drivers' decisions and negotiations under the circumstances.
    Therefore, we need to analyze and display individual driving performance via behavioral analysis of human factors using the relevant field metrics, which are commonly used~\cite{Barendswaard2019a, Barendswaard2019b}.
    In~\cite{HASTE2004}, \"Oslund et al. have recommended useful driving behavioral metrics to assess the driving performance of intelligent vehicle systems.
    We decided to adopt these metrics to further analyze driving performance.
    Among the feasible metrics, three of them are selected to analyze behavioral performance in addition to trajectory-based metrics: steering wheel reversal rates (SRRs), acceleration reversal rates (ARRs), and time-to-line crossing (TLC), measuring steering efficiency, pedaling efficiency, and safety behavior in terms of time margin, respectively.

    The reversal rate is used to analyze the efficiency of driving behavior.
    It is defined as the number of device reversals per minute, where the total number of reversals divided by the total driving time.
    A device reversal is counted up when the device angle rotates more than the specific angle gap after the direction of control is reversed.
    For SRRs, we calculate $SRR_{1^\circ}$, $SRR_{4^\circ}$, and $SRR_{8^\circ}$ according to the three gap sizes of 1$^\circ$, 4$^\circ$, and $8^\circ$ to show the efficiency of microscopic, mesoscopic, macroscopic steering movements, respectively.
    For ARRs, we select $ARR_{0.25^\circ}$, $ARR_{1^\circ}$, and $ARR_{2^\circ}$ considering that the accelerator pedal has a smaller control range than the steering wheel.

    We also examine the data with a safety metric called time-to-line-crossing~(TLC).
    It is defined as the available time margin before any part of the vehicle reaches the lane boundary condition assuming the vehicle remains in its current state.
    The TLC computation depends on the current vehicle state~\cite{Boer2016}.
    The vehicle is assumed to continue on a straight, counter-clockwise or clockwise rotational trajectory when $\theta_{s}\left[ k \right]=0^\circ$, $\theta_{s}\left[ k \right]<0^\circ$ and $\theta_{s}\left[ k \right]>0^\circ$, respectively.
    The time margin is then obtained by dividing the length of the assumed trajectory from the current location to the first intersection on the left/right boundary by the vehicle speed $v \left[ k \right]$.
    For our lane-following task on the first lane, we selected the centerline as the left boundary and the dividing line between the first and second lane as the right boundary, respectively.
    We separately compute $TLC_{fl}\left[ k \right]$ and $TLC_{fr}\left[ k \right]$ according to the relative positions of the left/right front wheels, and the actual safe time margin at $k$ will be $TLC\left[k\right]=\mathrm{min}\left(TLC_{fl}\left[k\right], TLC_{fr}\left[k\right] \right)$.
    We select $TLC_{<1}$ as representative safety performance.
    $TLC_{<1}$ is defined as the driving time with TLC less than a second divided by the total driving time, meaning the percentage of dangerous driving time.

\section{EXPERIMENT~I: MODELING VALIDITY}
\label{sec:exp1}

    Experiment~I aims to logically show that our NN models ($f_s$ and $f_a$) can capture expert driving skills.	

\subsection{Data Acquisition}
\label{sec:exp1:acquisition}

    The first goal of Experiment~I is to validate whether the models work in generalized environments.
    To this end, we designed longer, more complex paths by concatenating randomly-generated straight and curved segments, similar to \cite{Nechyba1998Thesis}.
    Each straight segment has one parameter, length $L$.
    Each curved segment has two parameters: the radius of curvature $R$ and the sweep angle $\phi$ (Fig. \ref{fig:driving_paths}).
    The parameters are randomly chosen from 100--150\,m ($L$ and $R$) and $\pm$45$^\circ$--135$^\circ$ ($\phi$; positive for left curves).
    A straight segment is followed by a left or right curve with an equal probability.
    A left (right) curve is followed by a straight segment with a probability of 0.4 and a right (left) curve with a probability of 0.6.
	The total length of each path is 4\,km.
	
\begin{figure}[!tb]
\centering
	\includegraphics[width=0.80\linewidth]{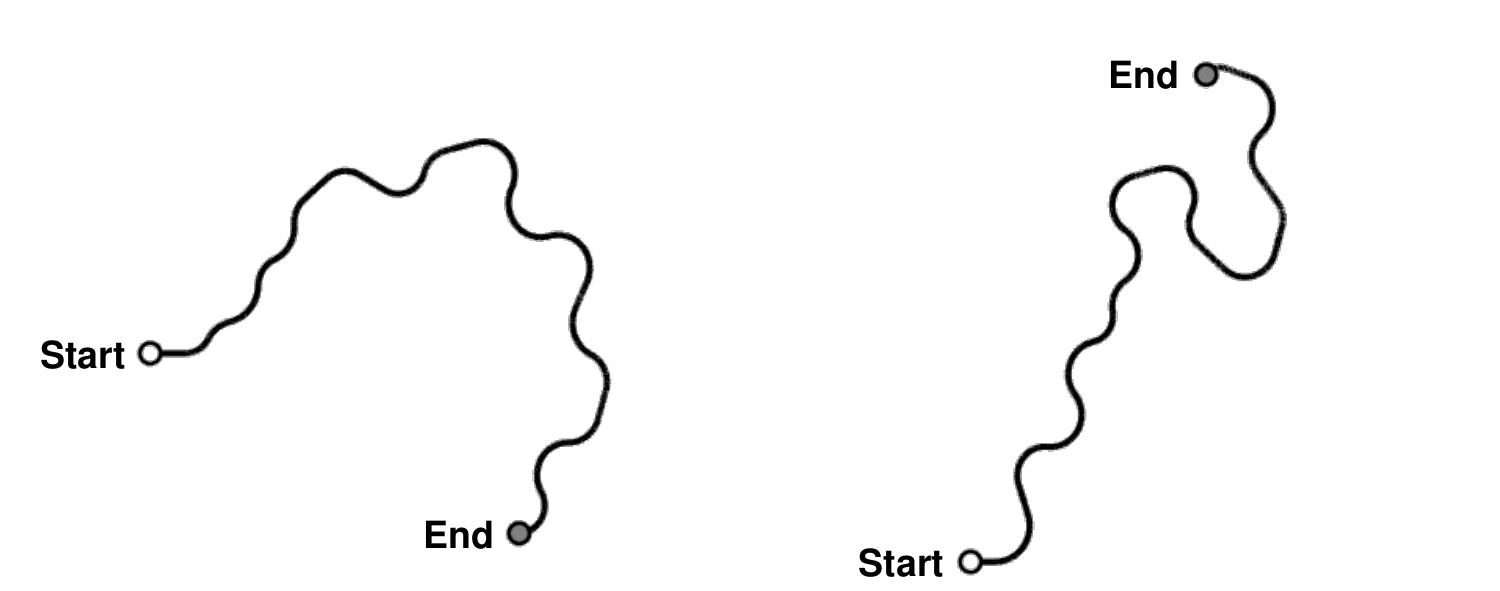}
	\caption{Driving paths used in Experiments I (left; 23 segments) and II (right; 22 segments). Total length: 4\,km.}
	\label{fig:exp_paths}
\end{figure}

    We randomly generated many paths and selected one representative path (Fig.~\ref{fig:exp_paths}, left).
    Compared to the short and simple paths used for the training of NNs (Section~\ref{sec:modeling:acquisition}), this path is considerably longer and more complex with arbitrary parameters of $L$, $R$, and $\phi$.
	
    The same five experienced (\textsf{EX}: $\mathsf{E_1}$--$\mathsf{E_5}$) and 18 novice drivers (\textsf{NO}: $\mathsf{N_1}$--$\mathsf{N_{18}}$; all males, age 18--28 years, M~22.8, SD~3.0) participated in collection of new driving data for the experiment.
    The novice participants either did not have driving licenses or had licenses but very little actual driving experience\footnote{Young individuals who had not owned or driven a car/motorcycle in the past two years.}.
    We controlled the novice drivers' gender and age as these are important factors for human motor skill studies.

    Only the novice participants had three practice trials in three 600-m short paths ($\phi=-90^\circ, 0^\circ, \mbox{and } 90^\circ$).
    They were instructed to drive the car close to the center of the first lane while maintaining the 60\,km/h speed.
	Then they proceeded to the main trial.
    Both the experienced and novice drivers completed the main trial on the 4-km long path (Fig.~\ref{fig:exp_paths}, left).

\subsection{Results}
\label{sec:exp1:results}

\begin{figure*}[!tb]
\centering
	\subfloat[Experienced \textsf{E4}. Right-sided figures demonstrate the magnified view at $t=$\,0--40\,(s).]
	{
        \includegraphics[width=0.95\textwidth]{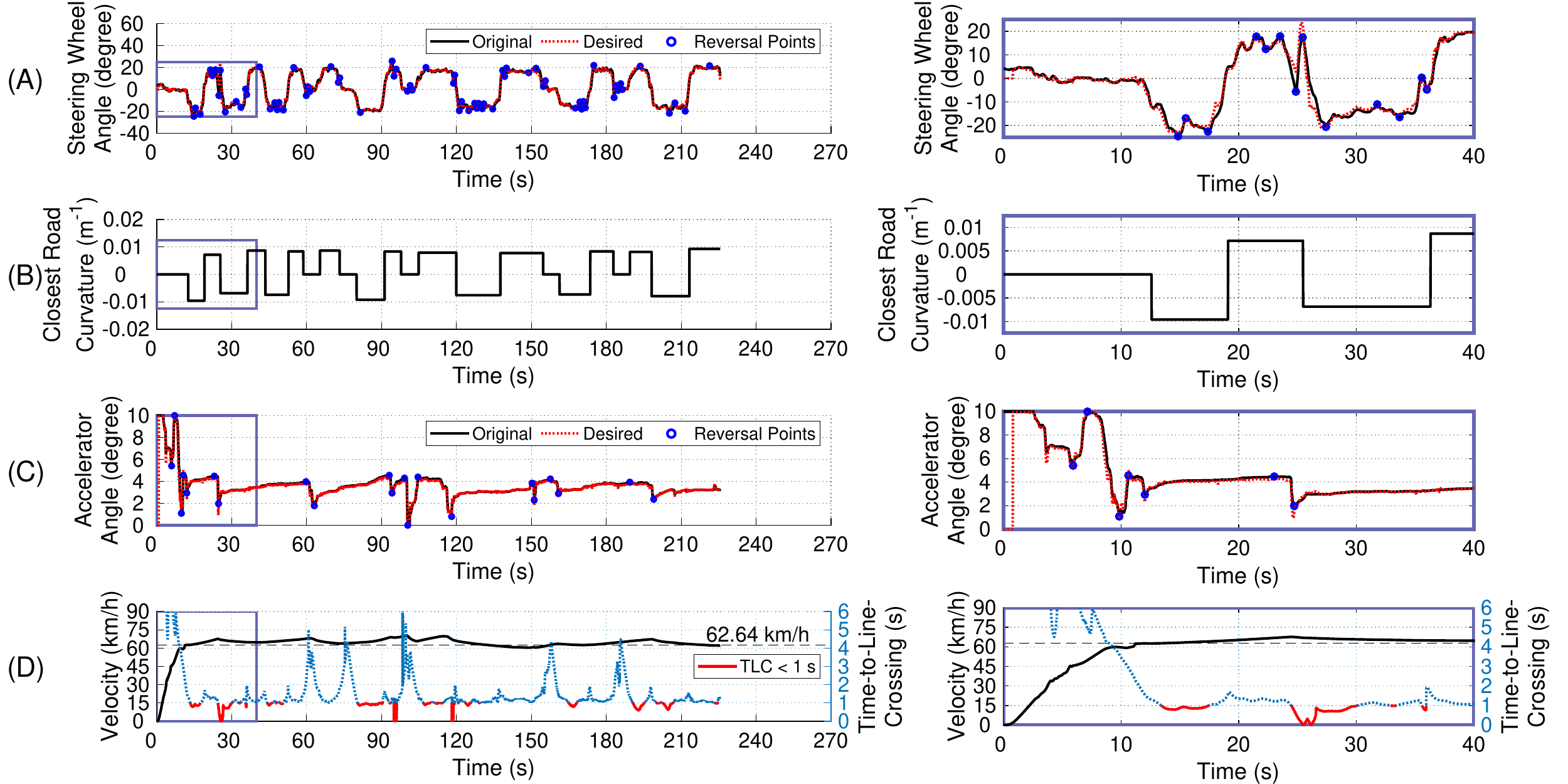}
        \label{fig:exp1_trajectory_exp}
    } \\
    \subfloat[Novice \textsf{N11}. Right-sided figures demonstrate the magnified view at $t=$\,0--40\,(s).]
	{
        \includegraphics[width=0.95\textwidth]{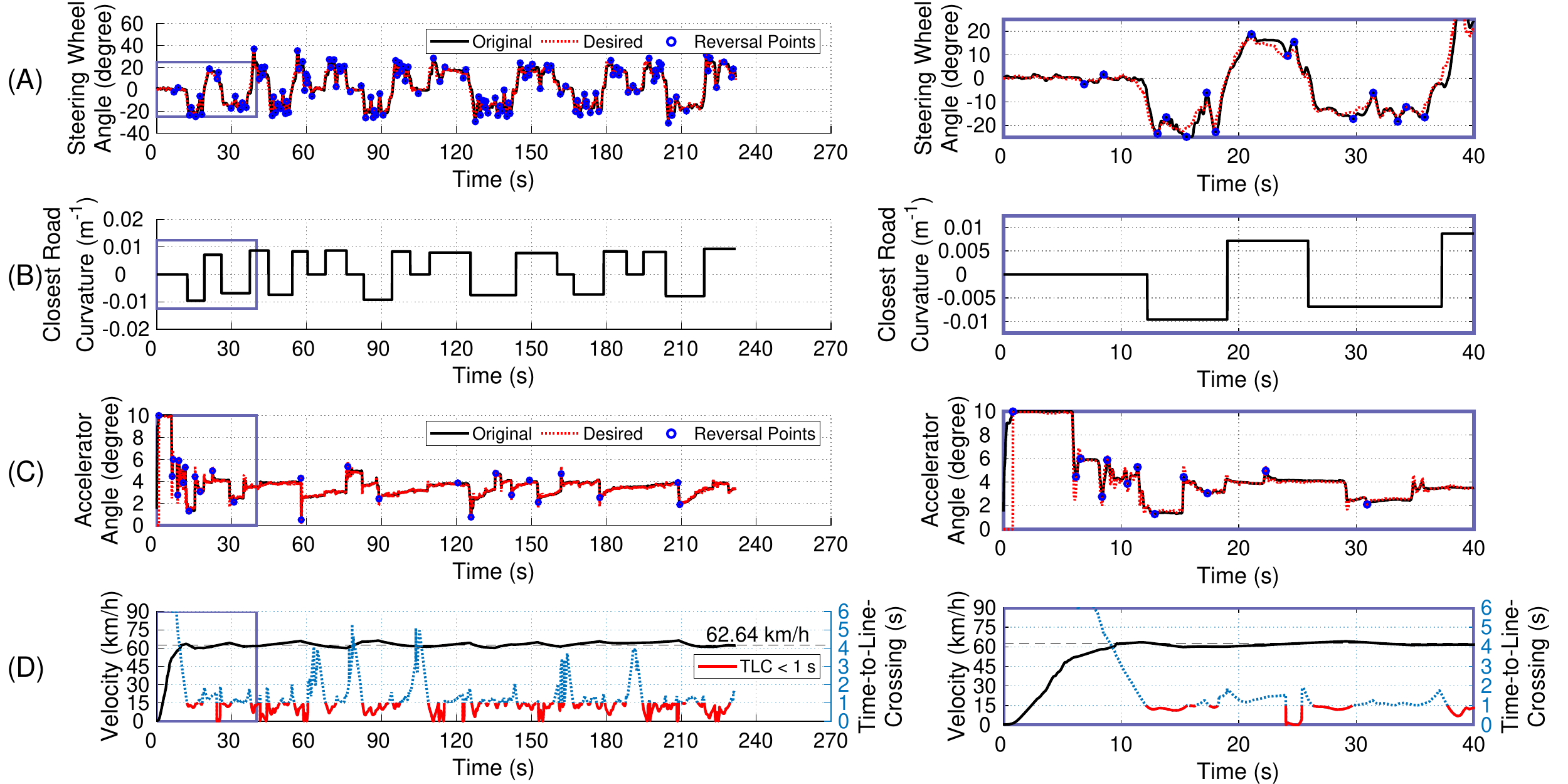}
        \label{fig:exp1_trajectory_nov}
    }
    \caption{Examples of the recorded driving data in Experiment~I. From the top of each driving data, (\textsf{A}) demonstrates the original steering wheel trajectory ($\theta_{s} \left( t \right)$; black, solid), steering wheel reversal points (blue markers; gap size $=4^\circ$), and the desired trajectory predicted by the NN ($\theta_{s, d} \left( t \right)$; red, dotted). (\textsf{B}) illustrates the curvature profile of the nearest road points from the current vehicle positions. (\textsf{C}) demonstrates the original accelerator trajectory ($\theta_{a} \left( t \right)$; black, solid), accelerator reversal points (blue markers; gap size $=1^\circ$) and the desired trajectories predicted by the NN ($\theta_{a, d} \left( t \right)$; red, dotted). (\textsf{D}) demonstrates $v\left(t\right)$ (black, solid) and $TLC\left(t\right)$ (blue, dotted); red colored parts indicate the TLC values less than 1\,s.}

    \label{fig:exp1_trajectory}
\end{figure*}

\begin{figure*}[!tb]
\centering
    \includegraphics[width=0.98\textwidth]{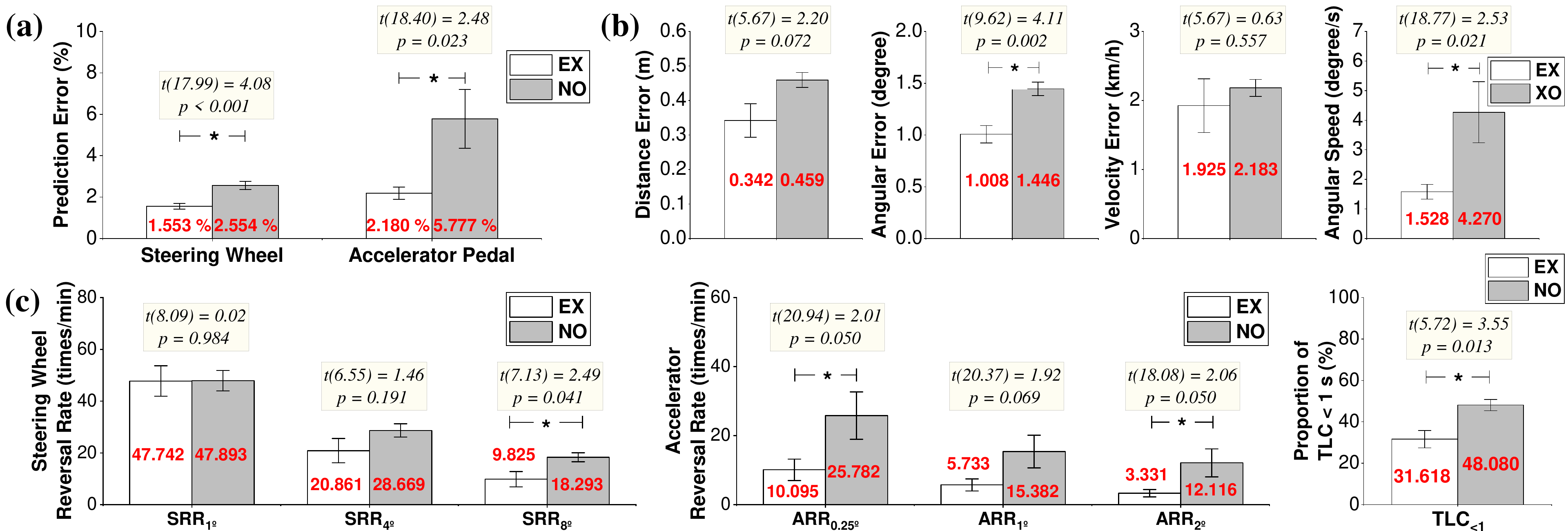}
    \caption{Means of all skill performance measures in Experiment~I. Error bars represent standard errors. The results of Welch's t-tests are also shown, with asterisks indicating significant differences ($\alpha = 0.05$). (a) Modeling performances ($\bar{E}_{s, p}$ and $\bar{E}_{a, p}$), (b) Trajectory-based performances ($E_{d}$, $E_{\delta}$, $E_{v}$, and $\Omega_{a}$), (c) Behavior performances ($SRR_{1^\circ}$, $SRR_{4^\circ}$, $SRR_{8^\circ}$, $ARR_{0.25^\circ}$, $ARR_{1^\circ}$, $ARR_{2^\circ}$, and $TLC_{<1}$).}
    \label{fig:exp1_all}
\end{figure*}

	Fig.~\ref{fig:exp1_trajectory} shows exemplar results for an experienced ($\mathsf{E_4}$) and a novice ($\mathsf{N_{11}}$) driver.
    They exhibited median performance for $\bar{E}_{s, p}$ and $\bar{E}_{a, p}$ in the respective groups.
    The experienced driver's trajectories appear to be more consistent with the desired trajectories generated by the NNs.

    The means of the 13 performance measures are shown in Fig.~\ref{fig:exp1_all}.
    We applied Welch's $t$-test (due to unequal sample sizes and unequal variances) to assess the effect of participant group (\textsf{EX} and \textsf{NO}) on each measure.
    The means and the results of Welch's $t$-tests are also shown in Fig.~\ref{fig:exp1_all}.
    All metrics showed that the \textsf{EX} group performed better driving than the \textsf{NO} group.
    The \textsf{EX} group showed significantly better performance than the \textsf{NO} group in $\bar{E}_{s, p}$, $\bar{E}_{a, p}$, $E_{\delta}$, and $\Omega_{a}$, and but not in $E_{d}$ and $E_{v}$.
    In the behavioral analysis, the \textsf{EX} group demonstrated significantly better performance than the \textsf{NO} group in $SRR_{8^\circ}$, $ARR_{0.25^\circ}$, and $TLC_{<1}$, but not in $SRR_{1^\circ}$, $SRR_{4^\circ}$, $ARR_{1^\circ}$, and $ARR_{2^\circ}$.

\subsection{Discussion}
\label{sec:exp1:discussion}

    There are three main findings we can acquire from the experiment.
    First, our models built with experienced drivers' actual driving data on short and simple paths still predict their driving actions effectively on longer, complicated paths.
    The normalized RMSEs of the \textsf{EX} group ($\bar{E}_{s, p} = 1.55\,\%$ and $\bar{E}_{a, p} = 2.18\,\%$) were similar to or less than the termination conditions of training ($\delta_{s} = 1.0\,\%$ and $\delta_{a} = 4.5\,\%$) (Section~\ref{sec:modeling:results}).

    Second, our models provide the skilled driving trajectories, the performance of which are similar to experienced drivers, and distinguished from superior to novice drivers.
    The \textsf{EX} group produced smaller normalized RMSEs ($\bar{E}_{s, p}$ and $\bar{E}_{a, p}$) than the \textsf{NO} group with statistical significance.
    The \textsf{EX} group also achieved better driving performance in every trajectory-based metric than \textsf{NO}, having statistically significant differences in $E_{\delta}$ and $\Omega_{a}$.
    In the behavioral analysis, the \textsf{EX} group was shown to have more efficient behavior in macroscopic steering movements (lower $SSR_{1^\circ}$) than the \textsf{NO} group.
    The \textsf{EX} group also had more efficient behavior in microscopic ($ARR_{0.25^\circ}$) and mesoscopic ($ARR_{2^\circ}$) pedaling movements.
    It is also observable that the \textsf{EX} group performed safer driving than the \textsf{NO} group, maintaining longer time with enough TLC margins.
    All these results indicate that the trained models can represent the specific driving skills of the experienced drivers, different from but better than those of the novice drivers.

    Third, however, it should be noted that the NN models can also induce human-like characteristics, which are shared between experienced and novice drivers.
    Although the \textsf{EX} group and the \textsf{NO} group demonstrated a large difference in $SRR_{8^\circ}$, the difference between the two groups was diminished as a smaller gap was applied.
    The efficiency of microscopic steering movements ($SRR_{1^\circ}$) of the \textsf{EX} group and that of the \textsf{NO} group was almost same.
    A similar result was already reported by a previous study~\cite{Greenshields1967}.
    Therefore, no difference in the efficiency of microscopic steering movements can be regarded as a common, humane attribute regardless of a driving skill level.
    This kind of natural human behavior could have been captured by our NN-based expert model.

\section{EXPERIMENT~II: HAPTIC ASSISTANCE}
\label{sec:exp2}

    Experiment~II aims to assess the overall validity of the framework, that integrates our expert skill models with performance-based haptic assistance.
    We compare three shared control methods, two of which are performance-based haptic assistance implemented in the form of haptic guidance.
    Both guidance methods utilize the performance error $\bm{e}_{\theta} = \bm{\theta} - \bm{\theta}_{d}$.
    However, our approach obtains $\bm{\theta}_{d}$ estimated by the NN models ($\hat{\bm{\theta}}$), whereas the conventional one deterministically formulates $\bm{\theta}_{d}$ with environmental variables.

\subsection{Shared Control Methods}
\label{sec:exp2:methods}

\begin{subsubsection}{No Guidance ($\mathsf{N}$)}

    A driver receives only realistic haptic feedback (Section~\ref{sec:simulation:torque}) while driving without any guidance.

\end{subsubsection}	
	
\begin{subsubsection}{Haptic Guidance with Neural Networks ($\mathsf{G}$)}

    A driver is assisted by performance-based guidance feedback, where the desired angle is computed by the expert skill model (NNs) in real-time.
    $\hat{\theta}_{s} \left[ k \right] $ and $\hat{\theta}_{a} \left[ k \right]$ estimated at 50\,Hz are upsampled and smoothed to $\hat{\theta}_{s} \left( t \right)$ and $\hat{\theta}_{a} \left( t \right)$ by moving average filters for 800-Hz torque feedback.
    The total steering torque feedback is
\begin{align}
\label{eq:TsAll}
	T_{s} & = T_{s, assist} + T_{s, stable},
\end{align}
    where $T_{s, stable}=-D_{stable} \dot{\theta}_{s}$ with increased viscosity but without Coulomb friction for stable feedback ($D_{stable} = 5D_{s}$).
    Aiming $\bm{\theta}_{d}=\hat{\bm{\theta}}$, the assistive torque $T_{s, assist}$ is computed using PID control such that
\begin{align}
\label{eq:Tsassist}
	T_{s, assist} \left( t \right)= -\Big( K_{pid} e_{s} \left( t \right) + I_{pid} \int_{t_{0}}^{t}{e_{s} \left( t' \right) dt'} + D_{pid} \dot{e}_{s} \Big),
\end{align}
\begin{align}
	e_{s} = \theta_{s} - \theta_{s, d} = \theta_{s} - \hat{\theta}_{s},
\end{align}
	where $t_{0}$ is the most recent time when $e_{s}$ was zero.
    $K_{pid} = 0.60$\,N$\cdot$m/degree, $I_{pid} = 0.12$\,N$\cdot$m$\cdot$s$^{-1}$/degree, and $D_{pid} = 0.06$\,N$\cdot$m$\cdot$s/degree.
    Those assistance gains in this PID control are adjustable.
    Therefore, this control logic can change the shared autonomy level between a driver and the driving system depending on the corresponding situation.

    In this experiment, we allowed the gains to be strong enough to let the novice drivers effectively follow the path by enhanced feedback and even finish driving with proper pedal maneuvers even without holding the steering wheel (i.e., like autonomous steering).
    However, to prevent the driver from the complete loss of human autonomy, the magnitude of gains is not excessively strong (mostly less than 10\,Nm; smaller than human steering torque capability~\cite{Forkenbrock2005}), so that the driver can also overpower the wheel to fine-tune its angle.
    We decided not to fix the driver's foot to the accelerator so that the driver can move the foot freely.
    The driver's foot and the accelerator cannot always be in full contact, so unlike the steering wheel, we use unidirectional torque feedback for pedaling.
    The total pedaling torque feedback is
\begin{align}
\label{eq:Tassist}
	T_{a} = T_{a, assist} + T_{a, spring} + T_{a, damping} + g(\theta_{a}).
\end{align}
    To aim $\bm{\theta}_{d}=\hat{\bm{\theta}}$,
\begin{align}
\label{eq:Taassist}
	T_{a, assist} \left( t \right) & =
	\begin{cases}
		0, & \: \text{if } \theta_{a} \left( t \right) < \hat{\theta}_{a} \left( t \right) \\
		-K_{a, max} \cdot e_{a} \left( t \right), & \: \text{if } \theta_{a} \left( t \right) \geq \hat{\theta}_{a} \left( t \right)
	\end{cases},
	\\
	e_{a} & = \theta_{a} - \theta_{a, d} =  \theta_{a} - \hat{\theta}_{a},
\end{align}
	which replaces $\theta_{a, max}$ in \eqref{eq:Tamax} with $\hat{\theta}_{a}$.
    With \eqref{eq:Taassist}, the accelerator presses the driver's foot upwards when the driver pushes the pedal deeper than $\hat{\theta}_{a}$.
    This control mechanism enables the driver to feel the desired pedal angle as an endpoint when pushing it.

\end{subsubsection}

\begin{subsubsection}{Conventional Haptic Guidance ($\mathsf{C}$)}

    A driver is assisted by the conventional performance-based guidance feedback, where the desired angles are determined from the vehicle configuration relative to the driving environment, i.e., the path.
    As such, $\mathsf{G}$ and $\mathsf{C}$ share the same torque feedback equations~\eqref{eq:Tsassist} and~\eqref{eq:Taassist}, but differ in how to determine $\bm{\theta}_{d}$.

    For steering, we compute $e_{s}$ ($= \theta_{s} - \theta_{s, d}$) as in the conventional predictive form of haptic steering guidance \cite{Marchal-Crespo2008, Lee2014}.
    The rationale is that guidance should be based on the future observation that a driver relies on to decide his/her current action.
    This method considers two error terms, a look-ahead direction error $e_{p}$ and the distance error $e_{d}$ (Fig.~\ref{fig:exp_error_diagram}).
	The desired angle $\theta_{s, d}$ is
\begin{align}
\label{eq:thetasd}
	\theta_{s, d} & = K_{p}e_{p} + K_{d}e_{d}.
\end{align}
    Using $K_{p}=7.65$, $K_{d}=1.00\,$degree/m, and the same PID gains of~\eqref{eq:Tsassist} in $\mathsf{G}$, the driver can also finish driving with proper pedaling even without holding the steering wheel.

    In the past, a few algorithms for pedaling feedback have been successfully implemented for specific goals such as safe car-following~\cite{Mulder2011}, safe curve initiation~\cite{Gruppelaar2018}, or fuel efficiency~\cite{Jamson2013}, but they are not yet generalized for performance-based haptic assistance.
    In Experiment~I, we realized that a novice driver could mistakenly accelerate due to little focus on pedals.
    Thus, we designed haptic feedback that only provides overspeed cues for drivers to keep the same velocity.
    We compute $e_{a}$ as follows:
\begin{align}
\label{eq:thetaad}
	e_{a} = \theta_{a} - \theta_{a, d} & =
	\begin{cases}
		\theta_{a} - \theta_{a, max}, & \: \text{if } v < v_{M}, \\
		\theta_{a} - \theta_{a, min}, & \: \text{if } v \geq v_{M}.
	\end{cases}
\end{align}
    This equation replaces $\theta_{a, d}$ with $\theta_{a, max}$ or $\theta_{a, min}$, concerning the criterion of overspeed\footnote{Exceeding 10\,\% from the target speed 60.0\,km/h.} $v_{M}=66.0$\,km/h.
    When the vehicle velocity exceeds $v_{M}$, a constant magnitude of torque generated from the instantaneous change of two constant endpoint positions from $\theta_{a, max}$ to $\theta_{a, min}$ is added to the feedback.
    As a result, the driver can feel immediate torque feedback from the foot as a warning sign of overspeed.

\end{subsubsection}

\subsection{Experimental Protocol}
\label{sec:exp2:protocol}
	
    We recruited the same 18 novice participants who participated in Experiment~I.
    To this end, we selected another randomly generated complex path (Fig.~\ref{fig:exp_paths}, right) to test three methods.
    Each participant completed three driving trials on the path with one of the three methods in each trial, in a within-subject design.
    The three methods were presented differently in a fully balanced order using all permutations across 18 participants.
    Each participant was paid KRW\,15,000 ($\simeq$~USD\,13) after the experiment.

    After each trial, the participants were asked to answer the following questions for both steering and pedaling feedback on a seven-point continuous scale:
        (1) (Effectiveness) Was the training effective for your driving?;
		(2) (Comfort) Was the training comfortable?;
		(3) (Fun) Was the training fun?; and
        (4) (Helpfulness) Would the training be helpful to improve your skill?

\begin{figure*}[!tb]
\centering
    \includegraphics[width=0.96\textwidth]{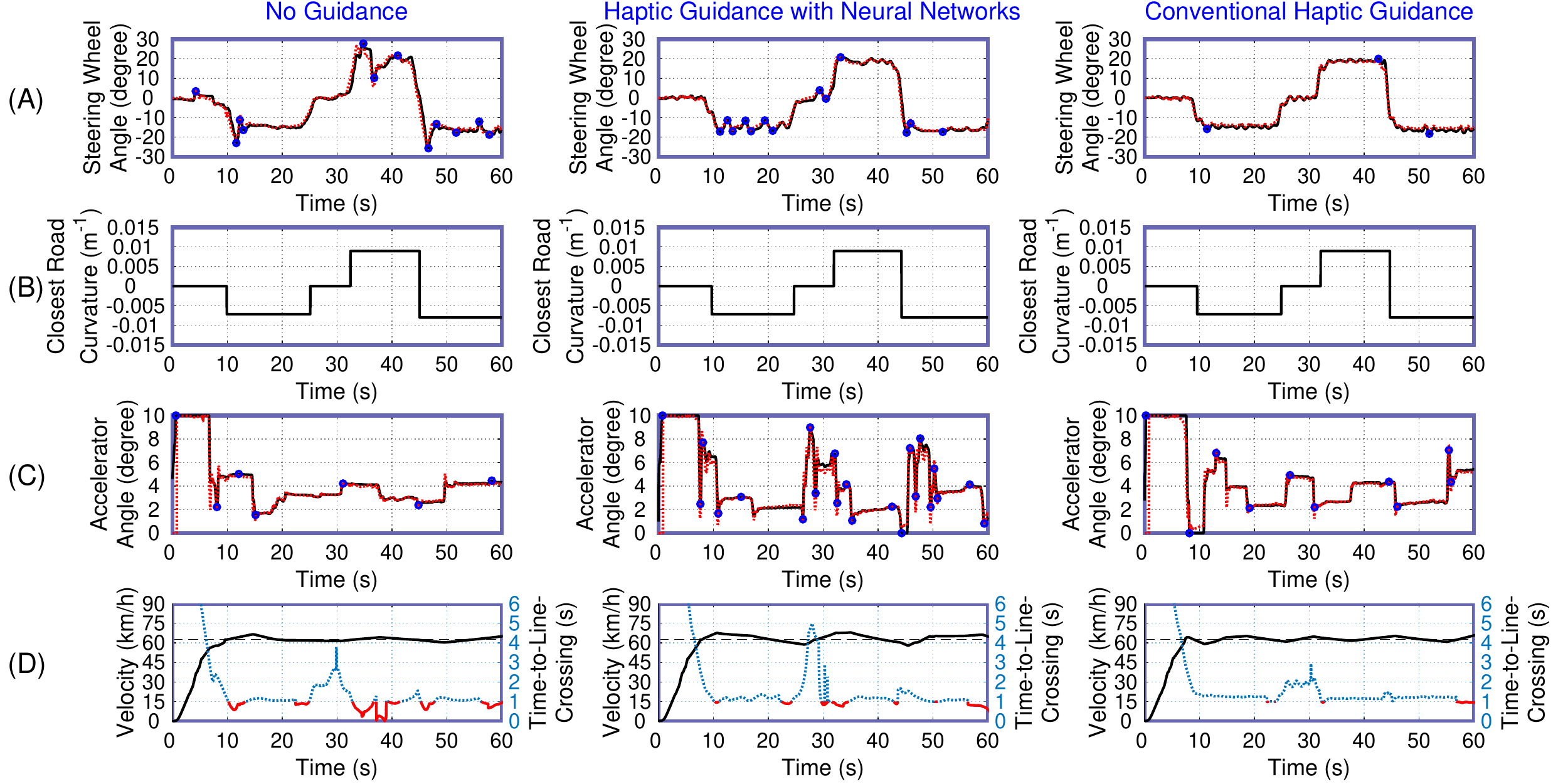}
    \caption{Examples of the recorded driving data in each method in Experiment~II (magnified at $t=$\,0--60\,(s)). From the top, (\textsf{A}) demonstrates the original steering wheel trajectories ($\theta_{s} \left( t \right)$; black, solid), steering wheel reversals (blue markers; gap size $=4^\circ$), and the desired trajectories predicted by the NN ($\theta_{s, d} \left( t \right)$; red, dotted). (\textsf{B}) illustrates the curvature profile of the nearest road points from the current vehicle positions. (\textsf{C}) demonstrates the original accelerator trajectory ($\theta_{a} \left( t \right)$; black, solid), accelerator reversals (blue markers; gap size $=1^\circ$), and the desired trajectories predicted by the NN ($\theta_{a, d} \left( t \right)$; red, dotted). (\textsf{D}) demonstrates $v\left(t\right)$ (black, solid) and $TLC\left(t\right)$ (blue, dotted); red colored parts indicate the TLC values less than 1\,s.}
    \label{fig:exp2_trajectory}
\end{figure*}

\subsection{Results}
\label{sec:exp2:results}

    Fig.~\ref{fig:exp2_trajectory} shows exemplar results of different assistance conditions.
    The resulted driving trajectories in every trial were analyzed using the same quantitative metrics used in Experiment~I.
    For statistical analysis, we applied a repeated-measures ANOVA with guidance method as the within-subject factor.
    Tukey's test was conducted as a post-hoc test for significant effects.

\begin{subsubsection}{Behavioral Similarity}

    We computed the normalized RMSEs as predictive errors, $\bar{E}_{s, p}$ and $\bar{E}_{a, p}$, for each trajectory.
    These metrics indicate how similar the participant's driving maneuvers are to the estimated outputs from the expert skill models.
    Their means are shown in Fig.~\ref{fig:exp2_all}a.

    The ascending order of $\bar{E}_{s, p}$ was $\mathsf{G} < \mathsf{C} < \mathsf{N}$.
    The data of $\bar{E}_{s, p}$ violated the assumption of sphericity (Mauchly's test, $\chi^{2} \left( 2 \right) = 29.04$, $p < 0.001$), and we applied the Greenhouse-Geisser correction ($\epsilon = 0.54$).
    Then we observed the significant effect of guidance method on $\bar{E}_{s, p}$ ($F(1.09, 18.51) = 34.27$, $p < 0.001$).
    The results of Tukey's test were $\mathsf{G} < \mathsf{N}$ and $\mathsf{C} < \mathsf{N}$ with statistical significance.

    The order of $\bar{E}_{a, p}$ was $\mathsf{N} < \mathsf{C} < \mathsf{G}$.
    The assumption of sphericity was not violated ($\chi^{2} \left( 2 \right) = 5.92$, $p = 0.052$), and there existed a significant effect of guidance method ($F(2, 34) = 5.55$, $p = 0.008$).
    Tukey's test showed that $\mathsf{N} < \mathsf{G}$ and $\mathsf{N} < \mathsf{C}$ with significance.

    The shared steering control in $\mathsf{G}$/$\mathsf{C}$ became similar to the predicted output from the expert skill model.
    When $\mathsf{G}$/$\mathsf{C}$ assisted the novice drivers, the drivers achieved small predictive errors similar to the termination condition of training ($\delta_{s}=1.0\,\%$).
    However, the shared pedaling control in $\mathsf{G}$/$\mathsf{C}$ was in discord with the output from the expert skill model.
    The drivers exhibited larger predictive errors of pedaling compared to the condition ($\delta_{a}=4.5\,\%$).

\end{subsubsection}

\begin{subsubsection}{Trajectory-based Skill Performance}

\begin{figure*} [!tb]
\centering
	\includegraphics[width=0.99\linewidth]{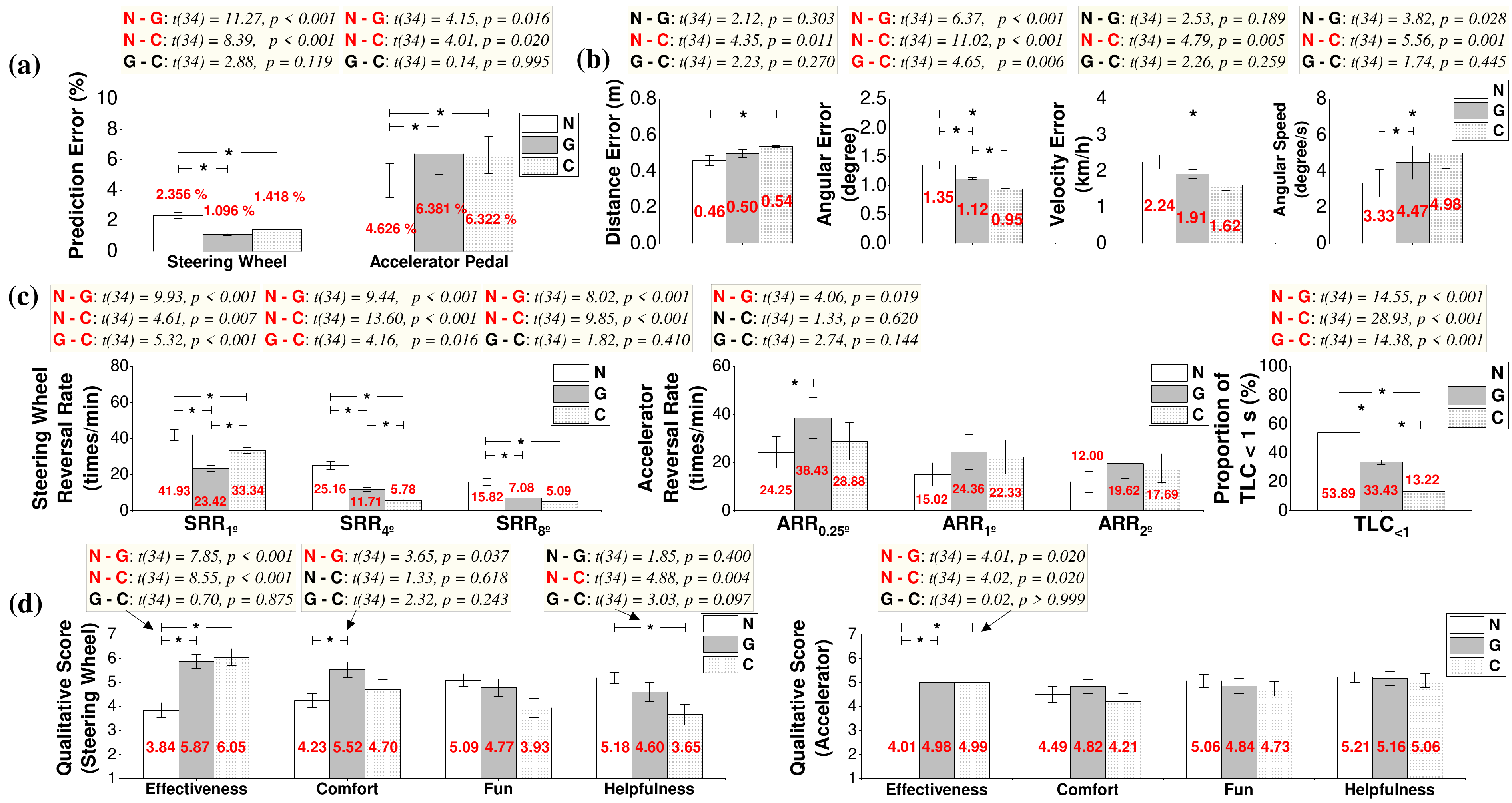}
	\caption{Means of all skill performance measures for each method in Experiment~II. Error bars represent standard errors. The results of Tukey's test are also shown in boxes, and asterisks (and red texts in each box) indicate statistically significant differences acquired from Tukey's test ($\alpha = 0.05$). (a) Modeling performances ($\bar{E}_{s, p}$ and $\bar{E}_{a, p}$), (b) Trajectory-based performances ($E_{d}$, $E_{\delta}$, $E_{v}$, and $\Omega_{a}$), (c) Behavior performances ($SRR_{1^\circ}$, $SRR_{4^\circ}$, $SRR_{8^\circ}$, $ARR_{0.25^\circ}$, $ARR_{1^\circ}$, $ARR_{2^\circ}$, and $TLC_{<1}$), (d) Subjective responses obtained from the questionnaire.}
\label{fig:exp2_all}
\end{figure*}

	The trajectory-based skill measures ($E_{d}$, $E_{\delta}$, $E_{v}$, and $\Omega_{a}$) were calculated for each driving trajectory.
    Their means are shown in Fig.~\ref{fig:exp2_all}b.

	The ranking of $E_{d}$ was
		$\mathsf{N} < \mathsf{G} < \mathsf{C}$,
	and the assumption of sphericity was not violated
		($\chi^{2} \left( 2 \right) = 1.74, p = 0.418$).
	There existed significant differences ($F(2, 34) = 4.73$, $p = 0.015$), and according to the post-hoc test,
		$\mathsf{N} < \mathsf{C}$.
	The ranking of $E_{\delta}$ was
        $\mathsf{C} < \mathsf{G} < \mathsf{N}$.
	Because the assumption of sphericity was violated
		($\chi^{2} \left( 2 \right) = 25.68, p < 0.001$),
	the Greenhouse-Geisser estimate of sphericity ($\epsilon = 0.56$) was used for correction.
	As a result, there existed significant differences ($F(1.11, 18.90) = 30.60$, $p < 0.001$), and the post-hoc test showed
		$\mathsf{G} < \mathsf{N}$,
		$\mathsf{C} < \mathsf{N}$,
	and
		$\mathsf{C} < \mathsf{G}$.
    The ranking of $E_{v}$ was
		$\mathsf{C} < \mathsf{G} < \mathsf{N}$.
	The assumption of sphericity was satisfied
		($\chi^{2} \left( 2 \right) = 1.24, p = 0.538$).
	Guidance method had significant effects ($F(2, 34) = 5.74$, $p = 0.007 $), and the post-hoc test showed that
		$\mathsf{C} < \mathsf{N}$.
	The ranking of $\Omega_{a}$ was
		$\mathsf{N} < \mathsf{G}  < \mathsf{C}$,
	and the assumption of sphericity was not violated
		($\chi^{2} \left( 2 \right) = 3.15, p = 0.207$).
	Guidance method was significant ($F(2, 34) = 8.08$, $p = 0.001$), and according to Tukey's test,
		$\mathsf{N} < \mathsf{G}$
	and
		$\mathsf{N} < \mathsf{C}$.

	In summary, $\mathsf{G}$ exhibited better performance than $\mathsf{N}$ in terms of $E_{\delta}$ but worse performance in $\Omega_{a}$.
	Between $\mathsf{C}$ and $\mathsf{N}$, $\mathsf{C}$ showed better performance in $E_{\delta}$ and $E_{v}$ but worse performance in $E_{d}$ and $\Omega_{a}$.
	Comparing the two guidance methods, $\mathsf{C}$ achieved better performance than $\mathsf{G}$ in $E_{\delta}$
	No other combinations showed significant differences.

\end{subsubsection}

\begin{subsubsection}{Driving Behavior Performance}

    The behavioral measures ($SRR_{1^\circ}$, $SRR_{4^\circ}$, $SRR_{8^\circ}$, $ARR_{0.25^\circ}$, $ARR_{1^\circ}$, $ARR_{2^\circ}$, and $TLC_{<1}$) of human factors in driving were calculated for each driving trajectory.
    The means are shown in Fig.~\ref{fig:exp2_all}c.

	The ranking of $SRR_{1^\circ}$ was
		$\mathsf{G} < \mathsf{C} < \mathsf{N}$,
	and the assumption of sphericity was violated
		($\chi^{2} \left( 2 \right) = 6.06, p = 0.048$).
    After the correction by the Greenhouse-Geisser estimate of sphericity ($\epsilon = 0.76$), there existed significant differences ($F(1.52, 25.85) = 24.70$, $p < 0.001$),
    and the post-hoc test showed
        $\mathsf{G} < \mathsf{N}$,
        $\mathsf{C} < \mathsf{N}$,
    and
		$\mathsf{G} < \mathsf{C}$.
	The ranking of $SRR_{4^\circ}$ was
        $\mathsf{C} < \mathsf{G} < \mathsf{N}$.
	The assumption of sphericity was violated
		($\chi^{2} \left( 2 \right) = 12.88, p = 0.002$),
	so the Greenhouse-Geisser estimate of sphericity ($\epsilon = 0.64$) was used for correction.
	There existed significant differences ($F(1.29, 21.90) = 48.55$, $p < 0.001$),
    and according to the post-hoc test,
		$\mathsf{G} < \mathsf{N}$,
		$\mathsf{C} < \mathsf{N}$,
	and
		$\mathsf{C} < \mathsf{G}$.
	The ranking of $SRR_{8^\circ}$ was also
        $\mathsf{C} < \mathsf{G} < \mathsf{N}$.
	Because the assumption of sphericity was violated as well
		($\chi^{2} \left( 2 \right) = 25.67, p < 0.001$),
	we applied the Greenhouse-Geisser estimate of sphericity ($\epsilon = 0.56$).
	In results, there existed significant differences ($F(1.11, 18.90) = 27.44$, $p < 0.001$).
    In the post-hoc test,
        $\mathsf{G} < \mathsf{N}$
    and
		$\mathsf{C} < \mathsf{N}$
    were shown to be statistically significant.

    The ranking of $ARR_{0.25^\circ}$ was
		$\mathsf{N} < \mathsf{C} < \mathsf{G}$,
	and the assumption of sphericity was not violated
		($\chi^{2} \left( 2 \right) = 1.72, p = 0.423$).
	There existed significant differences ($F(2, 34) = 4.30$, $p = 0.022$),
    and according to the post-hoc test,
		$\mathsf{N} < \mathsf{G}$.
	The rankings of $ARR_{1^\circ}$ and $ARR_{2^\circ}$ were also
        $\mathsf{N} < \mathsf{C} < \mathsf{G}$.
	Both satisfied the assumption of sphericity
		($\chi^{2} \left( 2 \right) = 3.54, p = 0.171$ and $\chi^{2} \left( 2 \right) = 1.79, p = 0.408$),
	but there existed no significant differences ($F(2, 34) = 2.42$, $p = 0.104$ and $F(2, 34) = 2.16$, $p = 0.130$).

    In $TLC_{<1}$, the ranking was
    	$\mathsf{C} < \mathsf{G} < \mathsf{N}$.
    The assumption of sphericity was not violated
		($\chi^{2} \left( 2 \right) = 1.74, p = 0.419$).
    There existed significant differences ($F(2, 34) = 209.30$, $p < 0.001$),
    and the post-hoc test showed that
		$\mathsf{G} < \mathsf{N}$,
		$\mathsf{C} < \mathsf{N}$,
	and
		$\mathsf{C} < \mathsf{G}$.

    In summary, both $\mathsf{G}$ and $\mathsf{C}$ showed better performance in all SRR metrics, $SRR_{1^\circ}$, $SRR_{4^\circ}$, and $SRR_{8^\circ}$.
    However, $\mathsf{G}$ showed the best performance in $SRR_{1^\circ}$, whereas $\mathsf{C}$ showed the best performance in $SRR_{4^\circ}$ and $SRR_{8^\circ}$, all with statistically significant differences.
    The ARR metrics showed that $\mathsf{G}$ and $\mathsf{C}$ were not efficient.
    In particular, the results of $ARR_{0.25^\circ}$ demonstrated that $\mathsf{G}$ significantly degrades the pedaling performance of $\mathsf{N}$.
    Both $\mathsf{G}$ and $\mathsf{C}$ showed that they could significantly help drivers drive safely with increased $TLC_{<1}$.
    This effect was more significant in $\mathsf{C}$ than in $\mathsf{G}$.

\end{subsubsection}

\begin{subsubsection}{Qualitative Results}

    We computed the mean scores for each subjective question (Fig.~\ref{fig:exp2_all}d).
    For the steering wheel, the ranking of effectiveness score was
        $\mathsf{N} < \mathsf{G} < \mathsf{C}$.
    The assumption of sphericity was violated
        ($\chi^{2} \left( 2 \right) = 10.87$, $p = 0.004$),
    and the Greenhouse-Geisser correction ($\epsilon = 0.67$) was applied.
    Significant differences were observed
        ($F(1.34, 22.77) = 22.53$, $p < 0.001$)
    and according to the post-hoc test,
        $\mathsf{N} < \mathsf{G}$
    and
        $\mathsf{N} < \mathsf{C}$ with significance.
    For comfort, the ranking was
        $\mathsf{N} < \mathsf{C} < \mathsf{G}$,
    and the assumption of sphericity was not violated
		($\chi^{2} \left( 2 \right) = 0.09$, $p = 0.957$).
    There existed a significant effect of guidance method ($F(2, 34) = 3.41$, $p = 0.045$),
    and the post-hoc test showed
		$\mathsf{N} < \mathsf{G}$.
    Both for fun and helpfulness, the ranking was
        $\mathsf{C} < \mathsf{G} < \mathsf{N}$.
    Because they violated the assumption of sphericity
		($\chi^{2} \left( 2 \right) = 7.41$, $p = 0.025$ and $\chi^{2} \left( 2 \right) = 6.28$, $p = 0.043$),
    the Greenhouse-Geisser correction was used
        ($\epsilon = 0.73$ and $0.76$).
    Significant differences were found in only helpfulness ($F(1.51, 25.67) = 6.06$, $p = 0.012$), and in the post-hoc test,
        $\mathsf{C} < \mathsf{N}$ with significance.

    For the accelerator pedal,
        $\mathsf{N} < \mathsf{G} < \mathsf{C}$
    in the effectiveness score. The assumption of sphericity was not violated
		($\chi^{2} \left( 2 \right) = 1.19$, $p = 0.550$).
    There existed significant differences ($F(2, 34) = 5.38$, $p = 0.009$), and the post-hoc test revealed that
		$\mathsf{N} < \mathsf{G}$,
    and
        $\mathsf{N} < \mathsf{C}$.
    The rankings of the comfort, fun, and helpfulness scores were
        $\mathsf{C} < \mathsf{N} < \mathsf{G}$,
        $\mathsf{C} < \mathsf{G} < \mathsf{N}$,
    and
        $\mathsf{C} < \mathsf{G} < \mathsf{N}$,
    respectively.
    However, no difference was not significantly observed in any of these three scores.

    In summary, for the steering feedback, significant differences were observed for effectiveness, comfort, and helpfulness.
    The participants reported that the two haptic guidance methods were more effective than $\mathsf{N}$.
    However, they reported that only $\mathsf{G}$ was more comfortable than $\mathsf{N}$, but $\mathsf{C}$ was less helpful than $\mathsf{N}$.
    The participants reported that the two haptic guidance methods were more effective for the pedaling feedback than $\mathsf{N}$.
    However, there existed no significant differences in comfort, fun, and helpfulness scores.

\end{subsubsection}

\subsection{Discussion}
\label{sec:exp2:discussion}

    For all behavior metrics, the performance of the novice drivers in Experiment~I and the performance of $\mathsf{N}$ in Experiment~II were similar, because the same novice drivers participated in both experiments.

    Two haptic guidance methods to enhance the steering performance of novice drivers were both effective.
    Both $\mathsf{G}$ and $\mathsf{C}$ could achieve smaller predictive errors ($\bar{E}_{s, p}$), smaller angular errors ($E_{\delta}$), lower reversal rates ($SRR_{1^\circ}$, $SRR_{4^\circ}$, $SRR_{8^\circ}$), and higher effectiveness scores, all of which are significantly discriminated from $\mathsf{N}$.
    In particular, $\mathsf{C}$ showed better performances than $\mathsf{G}$ in reducing angular errors and inducing efficient steering behavior in macroscopic movements with the lowest $SRR_{4^\circ}$, $SRR_{8^\circ}$.

    Interestingly, however, the proposed method $\mathsf{G}$ induced the most efficient steering behavior in microscopic movements with the lowest $SRR_{1^\circ}$, also with the best comfort score.
    In Experiment~I, both expert and novice drivers seem to have similar $SRR_{1^\circ}$, showing a natural human behavioral characteristic.
    However, in this experiment, $\mathsf{G}$ was shown to compensate the novice drivers' steering wheel reversals, resulting in more effective steering movements.
    The compensation effect was more significantly shown in $\mathsf{G}$ than $\mathsf{C}$.
    Therefore, this is a unique characteristic of $\mathsf{G}$, which is likely to be induced from the NN's output.

    However, all guidance cannot enhance the pedaling performance, and even sometimes can deteriorate the pedaling performance.
    $\mathsf{G}$ and $\mathsf{C}$ exhibited large predictive errors ($\bar{E}_{a, p}$), unstable pedal controls ($\Omega_{a}$), and low effectiveness scores, all of which were significantly discriminated from no assistance condition.
    $\mathsf{G}$ also had a significant detrimental effect on the efficiency of microscopic pedaling movements (high $ARR_{0.25^\circ}$), compared to the no assistance condition.

    Both $\mathsf{G}$ and $\mathsf{C}$ achieved lower $TLC_{<1}$ than $\mathsf{N}$, indicating that the novice drivers could drive with increased safe time margins under haptic guidance conditions.
    Although $\mathsf{G}$ showed higher $TLC_{<1}$ than $\mathsf{C}$, $\mathsf{G}$ could achieve similar $TLC_{<1}$ performance to that of the expert drivers in Experiment~I.
    It indicates that the novice drivers were able to drive as safely as the experts drive under the proposed method $\mathsf{G}$.

\section{GENERAL DISCUSSION}
\label{sec:discussion}

    Our driving task consists of two sub-tasks: lane keeping and speed control, and each sub-task focuses more on one of steering/pedaling skills, respectively.
    All limb movements in steering/pedaling actions for our task can be abstracted and generalized to human point-to-point movements.
    Point-to-point movements consist of two phases in the speed-accuracy tradeoff~\cite{Burdet1998, Corteville2007}.
    The first phase is a transfer motion, moving the body part close to the target point in a large displacement.
    A gross motor skill of the human within a feedforward mechanism is mainly relevant to perform this phase.
    The second is error-corrective movements added to the transfer motion when the body part reaches close enough to the target.
    A fine motor skill in a visuo-motor feedback loop is mainly required to execute these movements with faster and more accurately.

    The four trajectory-based metrics and the reversal rates with different gap sizes were separately designed to quantify the performance of two phased sub-skills (gross/fine motor skills).
    The angular error $E_{\delta}$ and the macroscopic steering wheel reversals quantify the performance in planning and initiating turns, which is reflected by the controllability of gross motor skills using arm joints.
    In contrast, the lateral error $E_{d}$ and the microscopic steering wheel reversals reflect the driving ability in immediate visuo-motor responses using hands and wrists.
    Similarly, the velocity error ($E_{v}$), the pedaling rate ($\Omega_{a}$), the macroscopic or microscopic accelerator reversal rates are also correlated to each performance of gross/fine motor ability in pedaling.

    Haptic guidance is considered adequate for gross motor skills~\cite{Marchal-Crespo2009}.
    Because the experts show superior performance in gross steering skills (Experiment~I), haptic guidance, which haptically demonstrates optimized skill execution, can be directly expected to enhance these motor skills (Experiment~II).
    However, it is still controversial whether haptic guidance can benefit fine motor skills (such as straight line-following) that require quick responses during execution~\cite{Marchal-Crespo2009,Lee2010}, so the enhancement of fine steering skills (Experiment~II) is not a common benefit of haptic guidance.
    Therefore, this effect cannot be interpreted as a general effect of haptic guidance.
    In this regard, it is highly likely a particular result of inherent motor adaptation~\cite{Bastian2008} that the novice drivers collaborate with the assistive haptic feedback on fine error-corrective movements of experts.
    Further investigation into motor adaption is required to clarify this phenomenon.

    The experts also show superior performance in fine pedaling skills (Experiment~I).
    However, our method (and also conventional haptic guidance) can enhance neither gross nor fine pedaling skills (Experiment~II).
    There exist several interpretations of this finding:
    First, haptic guidance usually helps to execute gross motor skills; haptic guidance may not have worked efficiently for this task.
    Second, the motor adaption of pedaling may not have occurred since the cognitive resources of novice drivers are usually limited~\cite{Engstrom2017}.
    Hence, it is difficult for them to interpret the assistance from both steering wheel feedback and pedal feedback at the same time.
    Third, some adjustments to the NN-based models should have been considered to induce better enhancement in pedaling skills.\looseness-1

    The third interpretation may infer that the NN-based models can be improved by taking human factors and practical driving skill characteristics into account based on the design flexibility of NN.
    We currently use the same training parameter $\tau = 10$ (0.2\,s) for both steering and pedaling models.
    However, the sensing accuracy and dexterity of lower limbs are often regarded as less than those of the hands~\cite{Velloso2015}.
    Thus, a longer time step ($\tau > 10$) with reduced movement bandwidth can be utilized for NN training to induce improved pedaling assistance.
    We may also leverage specified behavioral metrics for the input of NNs, such as safe time margins (TLC or time-to-extend-tangent-points (TETP)~\cite{Gruppelaar2018}).
    These metrics include human behavioral characteristics, so they can replace the environmental feature vectors ($\bm{d}$ or $\bm{z}$) that also manage hazardous driving conditions.\looseness-1

\section{CONCLUSIONS}
\label{sec:conclusion}

    This paper proposes a data-driven framework that consists of two parts: modeling expert driving skills from data to provide performance errors, and incorporating the model into performance-based haptic assistance to provide assistive haptic feedback to novice drivers.
    We developed a haptic driving simulator to collect expert driving data, and trained NNs with the data to build an expert skill model.
    Experiment~I validated that the expert models can provide an optimized reference for expert steering/pedaling actions, even when new and complex data is applied.
    In Experiment~II, we found that the performance-based haptic assistance utilizing our models can assist novice drivers' steering skills, but not pedaling skills.
    All of these results show that while our framework has some potentials to haptically assist driving skills, it is currently limited to only steering support.
    As a next step, we plan to conduct a long-term user study to investigate the educational effects of the framework.\looseness-1

    As a final remark, we note that our approach can be improved through integration with other models and algorithms.
    First, our framework can be regarded as an approach that applies Learning from Demonstration (LfD), widely used in robotics, to end users~\cite{Argall2009}.
    Thus, several machine-learning techniques based on human decision-making behavior, such as the Hidden Markov Model (HMM)~\cite{Nechyba1997}, would be a novel addition to our approach, especially for more complex driving tasks.
    Second, many other haptic assistance variations using performance error vectors can be candidates for applying the NN-based expert model.
    For example, progressive haptic guidance (or, guidance-as-needed) adjusts the amount of guidance feedback as the overall performance error changes~\cite{Marchal-Crespo2008}.
    Error amplification provides concurrent haptic feedback that amplifies performance errors~\cite{Emken2005}, and haptic disturbance extends error amplification by adding random and unpredictable noise~\cite{Lee2010}.
    Facilitation of such methods is also the intended direction in our future research.\looseness-1



\section*{ACKNOWLEDGMENT}

This research was supported by Ministry of Culture, Sports, and Tourism and Korea Creative Content Agency (Project Number: R2020040036).
The authors thank the reviewers for their constructive comments on their earlier manuscripts, especially about the suggestion of relevant performance metrics.
The authors also thank In Lee and Reza Haghighi Osgouei for their assistance in building the simulator.

\bibliographystyle{IEEEtran}
\bibliography{HMS_Driving}

\begin{thebibliography}{10}
\providecommand{\url}[1]{#1}
\csname url@samestyle\endcsname
\providecommand{\newblock}{\relax}
\providecommand{\bibinfo}[2]{#2}
\providecommand{\BIBentrySTDinterwordspacing}{\spaceskip=0pt\relax}
\providecommand{\BIBentryALTinterwordstretchfactor}{4}
\providecommand{\BIBentryALTinterwordspacing}{\spaceskip=\fontdimen2\font plus
\BIBentryALTinterwordstretchfactor\fontdimen3\font minus
  \fontdimen4\font\relax}
\providecommand{\BIBforeignlanguage}[2]{{%
\expandafter\ifx\csname l@#1\endcsname\relax
\typeout{** WARNING: IEEEtran.bst: No hyphenation pattern has been}%
\typeout{** loaded for the language `#1'. Using the pattern for}%
\typeout{** the default language instead.}%
\else
\language=\csname l@#1\endcsname
\fi
#2}}
\providecommand{\BIBdecl}{\relax}
\BIBdecl

\bibitem{Heuer2015}
H.~Heuer and J.~L{\"u}ttgen, ``Robot assistance of motor learning: {A}
  neuro-cognitive perspective,'' \emph{Neurosci. Biobehav. Rev.}, vol.~56, pp.
  222--240, 2015.

\bibitem{Gaffary2018}
Y.~Gaffary and A.~L{\'e}cuyer, ``The use of haptic and tactile information in
  the car to improve driving safety: {A} review of current technologies,''
  \emph{Frontiers in ICT}, vol.~5, p.~5, 2018.

\bibitem{Steele2001}
M.~Steele and R.~B. Gillespie, ``Shared control between human and machine:
  Using a haptic steering wheel to aid in land vehicle guidance,'' \emph{Proc.
  Hum. Factors Ergon. Soc. Annu. Meet.}, vol.~45, no.~23, pp. 1671--1675, 2001.

\bibitem{Griffiths2005}
P.~G. Griffiths and R.~B. Gillespie, ``Sharing control between humans and
  automation using haptic interface: primary and secondary task performance
  benefits,'' \emph{Hum. Factors}, vol.~47, no.~3, pp. 574--590, 2005.

\bibitem{Forsyth2006}
B.~A.~C. Forsyth and K.~E. MacLean, ``Predictive haptic guidance: {I}ntelligent
  user assistance for the control of dynamic tasks,'' \emph{{IEEE} Trans.
  Visual Comput. Graphics}, vol.~12, no.~1, pp. 103--113, 2006.

\bibitem{Saleh2013}
L.~Saleh, P.~Chevrel, F.~Claveau, J.-F. Lafay, and F.~Mars, ``Shared steering
  control between a driver and an automation: {S}tability in the presence of
  driver behavior uncertainty,'' \emph{IEEE Trans. Intell. Transp. Syst.},
  vol.~14, no.~2, pp. 974--983, 2013.

\bibitem{Abbink2011}
D.~A. Abbink, M.~Mulder, F.~C. Van~der Helm, and E.~Boer, ``Measuring
  neuromuscular control dynamics during car following with continuous haptic
  feedback,'' \emph{{IEEE} Trans. Syst. Man Cybern. B Cybern.}, vol.~41, no.~5,
  pp. 1239--1249, 2011.

\bibitem{Mulder2011}
M.~Mulder, D.~A. Abbink, M.~M. van Paassen, and M.~Mulder, ``Design of a haptic
  gas pedal for active car-following support,'' \emph{{IEEE} Trans. Intell.
  Transp. Syst.}, vol.~12, no.~1, pp. 268--279, 2011.

\bibitem{Jamson2013}
A.~H. Jamson, D.~L. Hibberd, and N.~Merat, ``The design of haptic gas pedal
  feedback to support eco-driving,'' in \emph{Proc. {Driving Assessment}},
  2013, pp. 264--270.

\bibitem{Omalley2006}
M.~K. O'Malley, A.~Gupta, M.~Gen, and Y.~Li, ``Shared control in haptic systems
  for performance enhancement and training,'' \emph{J. Dyn. Syst. Meas.
  Contr.}, vol. 128, no.~1, pp. 75--85, 2006.

\bibitem{Abbink2012}
D.~A. Abbink, M.~Mulder, and E.~R. Boer, ``Haptic shared control: {S}moothly
  shifting control authority?'' \emph{Cogn. Tech. Work}, vol.~14, no.~1, pp.
  19--28, 2012.

\bibitem{Gillespie1998}
R.~B. Gillespie, M.~S. O'Modhrain, P.~Tang, D.~Zaretzky, and C.~Pham, ``The
  virtual teacher,'' in \emph{Proc. {ASME DSC}}, 1998, pp. 171--178.

\bibitem{Mulder2012}
M.~Mulder, D.~A. Abbink, and E.~R. Boer, ``Sharing control with haptics:
  {S}eamless driver support from manual to automatic control,'' \emph{Hum.
  Factors}, vol.~54, no.~5, pp. 786--798, 2012.

\bibitem{Petermeijer2015}
S.~M. Petermeijer, D.~A. Abbink, M.~Mulder, and J.~C. de~Winter, ``The effect
  of haptic support systems on driver performance: {A} literature survey,''
  \emph{IEEE Trans. Haptics}, vol.~8, no.~4, pp. 467--479, 2015.

\bibitem{Sigrist2013}
R.~Sigrist, G.~Rauter, R.~Riener, and P.~Wolf, ``Augmented visual, auditory,
  haptic, and multimodal feedback in motor learning: a review,'' \emph{Psychon.
  Bull. Rev.}, vol.~20, no.~1, pp. 21--53, 2013.

\bibitem{Marchal-Crespo2008}
L.~Marchal-Crespo and D.~J. Reinkensmeyer, ``Haptic guidance can enhance motor
  learning of a steering task,'' \emph{J. Mot. Behav.}, vol.~40, no.~6, pp.
  545--556, 2008.

\bibitem{Marchal-Crespo2009}
L.~Marchal-Crespo, S.~McHughen, S.~C. Cramer, and D.~J. Reinkensmeyer, ``The
  effect of haptic guidance, aging, and initial skill level on motor learning
  of a steering task,'' \emph{Exp. Brain Res.}, vol. 201, no.~2, pp. 209--220,
  2009.

\bibitem{Lee2014}
H.~Lee and S.~Choi, ``Combining haptic guidance and haptic disturbance: {A}n
  initial study of hybrid haptic assistance for virtual steering task,'' in
  \emph{Proc. {IEEE HAPTICS}}, 2014, pp. 159--165.

\bibitem{Aksjonov2018}
A.~Aksjonov, P.~Nedoma, V.~Vodovozov, E.~Petlenkov, and M.~Herrmann, ``A novel
  driver performance model based on machine learning,''
  \emph{IFAC-PapersOnLine}, vol.~51, no.~9, pp. 267--272, 2018.

\bibitem{Abou2020}
Z.~E. Abou~Elassad, H.~Mousannif, H.~Al~Moatassime, and A.~Karkouch, ``The
  application of machine learning techniques for driving behavior analysis: A
  conceptual framework and a systematic literature review,'' \emph{Eng. Appl.
  Artif. Intell.}, vol.~87, p. 103312, 2020.

\bibitem{Nechyba1996}
M.~C. Nechyba and Y.~Xu, ``On the fidelity of human skill models,'' in
  \emph{Proc. {IEEE ICRA}}, vol.~3, 1996, pp. 2688--2693.

\bibitem{Nechyba1997}
------, ``Human control strategy: {A}bstraction, verification, and
  replication,'' \emph{IEEE Control Syst. Mag.}, vol.~17, no.~5, pp. 48--61,
  1997.

\bibitem{Nechyba1998}
------, ``Stochastic similarity for validating human control strategy models,''
  \emph{IEEE Trans. Robot. Autom.}, vol.~14, no.~3, pp. 437--451, 1998.

\bibitem{Lin2005}
Y.~Lin, P.~Tang, W.~Zhang, and Q.~Yu, ``Artificial neural network modelling of
  driver handling behaviour in a driver-vehicle-environment system,''
  \emph{Int. J. Veh. Des.}, vol.~37, no.~1, pp. 24--45, 2005.

\bibitem{Garimella2017}
G.~Garimella, J.~Funke, C.~Wang, and M.~Kobilarov, ``Neural network modeling
  for steering control of an autonomous vehicle,'' in \emph{Proc. {IEEE/RSJ
  IROS}}, 2017, pp. 2609--2615.

\bibitem{VPP}
Vehicle Physics Pro (VPP). http://www.vehiclephysics.com/.

\bibitem{AutomobileCatalog}
Automobile-Catalog. https://www.automobile-catalog.com/.

\bibitem{hiraoka2008}
T.~Hiraoka, O.~Nishihara, and H.~Kumamoto, ``Steering reactive torque
  presentation method for a steer-by-wire vehicle,'' \emph{Rev. Automot. Eng.},
  vol.~29, pp. 287--294, 2008.

\bibitem{Marchal-Crespo2009Review}
L.~Marchal-Crespo and D.~J. Reinkensmeyer, ``Review of control strategies for
  robotic movement training after neurologic injury,'' \emph{J. NeuroEng.
  Rehabil.}, vol.~6, no.~1, p.~20, 2009.

\bibitem{Nechyba1995}
M.~C. Nechyba and Y.~Xu, ``Human skill transfer: {N}eural networks as learners
  and teachers,'' in \emph{Proc. {IEEE/RSJ IROS}}, vol.~3, 1995, pp. 314--319.

\bibitem{Narendra1990}
K.~S. Narendra and K.~Parthasarathy, ``Identification and control of dynamical
  systems using neural networks,'' \emph{IEEE Trans. Neural Netw.}, vol.~1,
  no.~1, pp. 4--27, 1990.

\bibitem{HCM2010}
\emph{Highway Capacity Manual 2010}, 5th~ed., Transporation Research Board of
  the National Academies of Science, Washington, DC, USA, 2010.

\bibitem{Neudorff2016}
L.~G. Neudorff, P.~Jenior, R.~G. Dowling, and B.~L. Nevers, ``Use of narrow
  lanes and narrow shoulders on freeways: {A} primer on experiences, current
  practice, and implementation considerations,'' U.S. Department of
  Transporation, Federal Highway Administration (FHWA), Office of Operations,
  Tech. Rep. FHWA HOP-16-060, 2016.

\bibitem{Brooks1990}
T.~L. Brooks, ``Telerobotic response requirements,'' in \emph{Proc. {IEEE
  SMC}}, 1990, pp. 113--120.

\bibitem{Barendswaard2019a}
S.~Barendswaard, D.~M. Pool, E.~R. Boer, and D.~A. Abbink, ``A classification
  method for driver trajectories during curve-negotiation,'' in \emph{Proc.
  IEEE SMC}, 2019, pp. 3729--3734.

\bibitem{Barendswaard2019b}
S.~Barendswaard, L.~van Breugel, B.~Schelfaut, J.~Sluijter, L.~Zuiker, D.~M.
  Pool, E.~R. Boer, and D.~A. Abbink, ``Effect of velocity and curve radius on
  driver steering behaviour before curve entry,'' in \emph{Proc. IEEE SMC},
  2019, pp. 3866--3871.

\bibitem{HASTE2004}
J.~{\"O}stlund, L.~Nilsson, O.~Carsten, N.~Merat, H.~Jamson, S.~Jamson,
  S.~Mouta, J.~Carvalhais, J.~Santos, V.~Anttila, H.~Sandberg, J.~Luoma,
  D.~de~Waard, K.~Brookhuis, E.~Johansson, J.~Engström, T.~Victor, J.~Harbluk,
  W.~Janssen, and R.~Brouwer, ``{HASTE} deliverable 2: {HMI} and safety-related
  driver performance,'' Human Machine Interface And the Safetyof Traffic in
  Europe (HASTE) Project, Tech. Rep. GRD1/2000/25361 S12.319626, 2004.

\bibitem{Boer2016}
E.~R. Boer, ``Satisficing curve negotiation: {E}xplaining drivers’ situated
  lateral position variability,'' \emph{IFAC-PapersOnLine}, vol.~49, no.~19,
  pp. 183--188, 2016.

\bibitem{Nechyba1998Thesis}
M.~C. Nechyba, ``Learning and validation of human control strategies,'' Ph.D.
  Thesis, The Robotics Institue, Carnegie Mellon University, 1998.

\bibitem{Greenshields1967}
B.~D. Greenshields and F.~N. Platt, ``Development of a method of predicting
  high-accident and high-violation drivers.'' \emph{J. Appl. Psychol.},
  vol.~51, no.~3, p. 205, 1967.

\bibitem{Forkenbrock2005}
G.~J. Forkenbrock and D.~Elsasser, ``An assessment of human driver steering
  capability,'' National Highway Traffic Safety Administration, Vehicle
  Research and Test Center, Tech. Rep. DOT HS 809 875, 2005.

\bibitem{Gruppelaar2018}
V.~Gruppelaar, R.~van Paassen, M.~Mulder, and D.~Abbink, ``A perceptually
  inspired driver model for speed control in curves,'' in \emph{Proc. IEEE
  SMC}, 2018, pp. 1257--1262.

\bibitem{Burdet1998}
E.~Burdet and T.~E. Milner, ``Quantization of human motions and learning of
  accurate movements,'' \emph{Biol. Cybern.}, vol.~78, no.~4, pp. 307--318,
  1998.

\bibitem{Corteville2007}
B.~Corteville, E.~Aertbelien, H.~Bruyninckx, J.~De~Schutter, and
  H.~Van~Brussel, ``Human-inspired robot assistant for fast point-to-point
  movements,'' in \emph{Proc. {IEEE ICRA}}, 2007, pp. 3639--3644.

\bibitem{Lee2010}
J.~Lee and S.~Choi, ``Effects of haptic guidance and disturbance on motor
  learning: {P}otential advantage of haptic disturbance,'' in \emph{Proc.
  {IEEE} {HAPTICS}}, 2010, pp. 335--342.

\bibitem{Bastian2008}
A.~J. Bastian, ``Understanding sensorimotor adaptation and learning for
  rehabilitation,'' \emph{Curr. Opin. Neurol.}, vol.~21, no.~6, pp. 628--633,
  2008.

\bibitem{Engstrom2017}
J.~Engstr{\"o}m, G.~Markkula, T.~Victor, and N.~Merat, ``Effects of cognitive
  load on driving performance: {T}he cognitive control hypothesis,'' \emph{Hum.
  Factors}, vol.~59, no.~5, pp. 734--764, 2017.

\bibitem{Velloso2015}
E.~Velloso, D.~Schmidt, J.~Alexander, H.~Gellersen, and A.~Bulling, ``The feet
  in human--computer interaction: {A} survey of foot-based interaction,''
  \emph{ACM Comput. Surv.}, vol.~48, no.~2, pp. 21:1--21:35, 2015.

\bibitem{Argall2009}
B.~D. Argall, S.~Chernova, M.~Veloso, and B.~Browning, ``A survey of robot
  learning from demonstration,'' \emph{Rob. Auton. Syst.}, vol.~57, no.~5, pp.
  469--483, 2009.

\bibitem{Emken2005}
J.~L. Emken and D.~J. Reinkensmeyer, ``Robot-enhanced motor learning:
  {A}ccelerating internal model formation during locomotion by transient
  dynamic amplification,'' \emph{{IEEE} Trans. Neural Syst. Rehabil. Eng.},
  vol.~13, no.~1, pp. 33--39, 2005.

\end{thebibliography}

\begin{IEEEbiography}[{\includegraphics[width=1in,height=1.25in,clip,keepaspectratio]{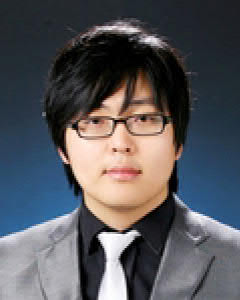}}]{Hojin Lee}
is a postdoctoral researcher in Haptic Intelligence Department at Max Planck Institute for Intelligent Systems, Stuttgart, Germany.
He received the B.Sc. and Ph.D. degrees in Computer Science and Engineering from Pohang University of Science and Technology (POSTECH) in 2010 and 2019, respectively.
His research interests lie on haptics, virtual reality, human-computer interaction, motor learning, cognitive psychology with human factors, and other human-machine technologies.
He is a member of the IEEE.
\end{IEEEbiography}

\vfill

\begin{IEEEbiography}[{\includegraphics[width=1in,height=1.25in,clip,keepaspectratio]{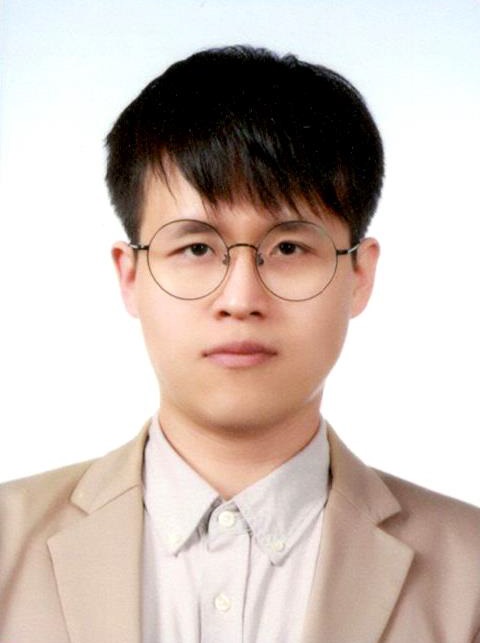}}]{Hyoungkyun Kim}
is a staff engineer in Samsung Research, Samsung Electronics Co., Ltd, Seoul, South Korea.
He received the B.Sc. and Ph.D. degrees in Mechanical Engineering from Pohang University of Science and Technology (POSTECH) in 2010 and 2018, respectively.
His research interests lie on medical robot, tactile sensing and other haptic technologies.
\end{IEEEbiography}

\vfill

\begin{IEEEbiography}[{\includegraphics[width=1in,height=1.25in,clip,keepaspectratio]{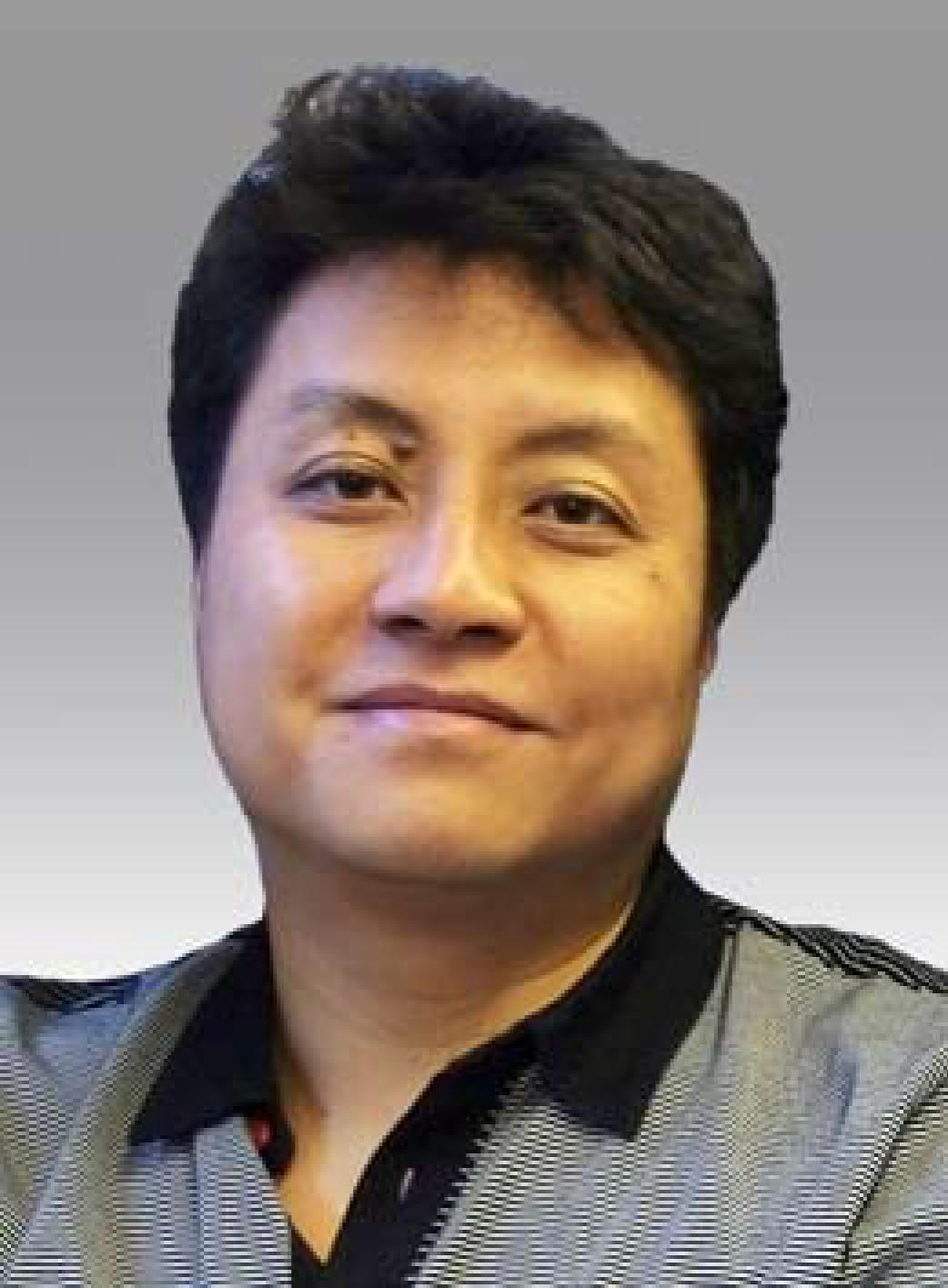}}]{Seungmoon Choi}
is a professor of Computer Science and Engineering at Pohang University of Science and Technology (POSTECH), Pohang, South Korea.
He received the B.Sc. and M.Sc. degrees in Control and Instrumentation Engineering from Seoul National University in 1995 and 1997, respectively, and the Ph.D. degree in Electrical and Computer Engineering from Purdue University in 2003.
His research interests lie on haptic rendering and perception, both in kinesthetic and tactile aspects.
He is a senior member of the IEEE.
\end{IEEEbiography}

\vfill

\end{document}